\pdfoutput = 1
\documentclass[twocolumn]{article}
\usepackage[utf8]{inputenc}
\usepackage[dvipdfmx]{graphicx}
\usepackage{siunitx}
\usepackage{authblk}
\usepackage{hyperref}

\usepackage{amsmath}
\usepackage{amssymb}
\usepackage{multirow}
\usepackage{subfig}

\hypersetup{
    colorlinks=true,
    linkcolor=blue,
    filecolor=magenta,      
    urlcolor=cyan,
    }

\usepackage[margin=2.5cm]{geometry}

\title{\vspace{-15mm} Image features of a splashing drop on a solid surface \\ extracted using a feedforward neural network}

\author[1]{\small{Jingzu Yee}}
\author[1]{Akinori Yamanaka}
\author[1,2]{Yoshiyuki Tagawa}

\affil[1]{Department of Mechanical Systems Engineering, Tokyo University of Agriculture and Technology, Koganei, Tokyo 184-8588, Japan
}
\affil[2]{Institute of Global Innovation Research, Tokyo University of Agriculture and Technology, Koganei, Tokyo 184-8588, Japan
}

\date{}

\begin{document}

\twocolumn[

\maketitle

\vspace{-10mm}

\paragraph{Abstract}
This article reports nonintuitive characteristic of a splashing drop on a solid surface discovered through extracting image features using a feedforward neural network (FNN).
Ethanol of area-equivalent radius about 1.29~mm was dropped from impact heights ranging from 4 to 60~cm (splashing threshold 20~cm) and impacted on a hydrophilic surface.
The images captured when half of the drop impacted the surface were labeled according to their outcome, splashing or nonsplashing, and were used to train an FNN. A classification accuracy $\geq 96\%$ was achieved.
To extract the image features identified by the FNN for classification, the weight matrix of the trained FNN for identifying splashing drops was visualized.
Remarkably, the visualization showed that the trained FNN identified the contour height of the main body of the impacting drop as an important characteristic differentiating between splashing and nonsplashing drops, which has not been reported in previous studies.
This feature was found throughout the impact, even when one and three-quarters of the drop impacted the surface.
To confirm the importance of this image feature, the FNN was retrained to classify using only the main body without checking for the presence of ejected secondary droplets.
The accuracy was still $\geq 82\%$, confirming that the contour height is an important feature distinguishing splashing from nonsplashing drops.
Several aspects of drop impact are analyzed and discussed with the aim of identifying the possible mechanism underlying the difference in contour height between splashing and nonsplashing drops.

\vspace{20pt}

]

\section{\label{sec:intro}Introduction}

The impact of a liquid drop on a solid surface is an important phenomenon that occurs frequently both in nature and in industry \cite{josserand2016drop, yarin2006drop}.
Many different physical properties are involved in this phenomenon, such as the properties of the liquid drop (e.g., its velocity, surface tension, and viscosity), the conditions of the solid surface (e.g., its temperature, roughness, and stiffness), and the ambient conditions (e.g., temperature, pressure, and humidity) \cite{josserand2016drop, yarin2006drop, du2021initial, sahoo2021collisional, yokoyama2021droplet, usawa2021large, hatakenaka2019magic, wang2017wetting, kim2014drop}.
Thus, there are various possible outcomes when a drop impacts on a solid surface \cite{yarin2006drop, rioboo2001outcomes, zhao2018splashing, josserand2016drop}.

A major outcome is splashing, which occurs when the impacting drop breaks up and ejects secondary droplets \cite{gordillo2019note,riboux2014experiments,riboux2017boundary,burzynski2020splashing}.
By contrast, a nonsplashing drop just spreads over the surface until it reaches a maximum radius \cite{gordillo2019theory,clanet2004maximal, baroudi2020effect}.

The study of drop impact has evolved enormously from the time when only a few stages of this high-speed phenomenon could be observed to the recent advent of high-speed videography that has enabled the observation of microdrop impact at a rate of a frame every 100~ns \cite{worthington1877xxviii, thoroddsen2008high, visser2015dynamics, visser2012microdroplet}.
Nevertheless, observation and study of the drop impact still rely heavily on frame-by-frame inspection with human eyes.
However, owing to the complex nature of the phenomenon, many important but nonintuitive characteristics could possibly be missed when observation is with the naked eye alone.

Fortunately, the tremendous advances in machine learning techniques brought about by the recent boom in artificial intelligence (AI) seem to have provided an answer.
Artificial neural networks (ANNs), supervised machine learning algorithms inspired by biological neural networks, have been widely utilized and have proven accurate for various classification and prediction tasks \cite{rosenblatt1958perceptron, hornik1989multilayer, hornik1991approximation, chen2019design}.
For example, the ability of a deep convolutional neural network (CNN) to accurately classify images has been widely exploited in search engines, face recognition, and cancer diagnosis, among many other tasks \cite{krizhevsky2012imagenet, he2016deep, esteva2017dermatologist, lawrence1997face}.
Already in the field of fluid mechanics \cite{brunton2020machine}, ANNs have been utilized for various purposes \cite{gundersen2021semi, li2021efficient, pawar2021physics, xu2021deep, erichson2020shallow, li2020machine, pawar2020interface, peng2020unsteady}, such as bubble pattern recognition \cite{poletaev2016artificial}, turbulence modeling \cite{nakamura2021convolutional, yin2020feature, ling2016reynolds}, and classification of vortex wakes \cite{colvert2018classifying}.

Despite their proven prediction and classification accuracy, machine learning models are often too complicated and usually function as black boxes, with the designers being unable to explain the underlying reasoning that leads to a specific decision \cite{arrieta2020explainable, adadi2018peeking}.
However, previous studies have shown that a simple and interpretable model such as a feedforward neural network (FNN) can achieve high performance when trained with highly similar and high-quality data even if the amount of training data is limited \cite{colvert2018classifying,erichson2020shallow,bright2013compressive}.

Therefore, this study aims to unveil important but nonintuitive characteristics of the splashing of a drop on a solid surface by extracting the image features that a well-trained and highly accurate FNN model uses to classify images of splashing and nonsplashing drops during their impact.
In Sec.~\ref{sec:method}, the methodology of the study, including data collection, data preparation, and image classification using an FNN, is explained in detail.
In Sec.~\ref{sec:result}, the results, including the classification performance and an analysis of the classification process of the trained FNN, are presented and discussed.

\section{\label{sec:method}Methodology}

The methodology of this study can be summarized as follows.
A drop impact experiment is performed to capture high-speed videos of impacts of splashing and nonsplashing drops on a solid surface using a high-speed camera (Sec.~\ref{sec:drop_imp_exp}).
To ensure high similarity and quality of the images, digital image processing is performed using an in-house MATLAB code to extract the desired frames and crop away the unnecessary image background (Sec.~\ref{sec:dip}).
Next, the processed images are segmented (Sec.~\ref{sec:data_segment}) to train, validate, and test an FNN until high accuracy is achieved (Sec.~\ref{sec:fnn}).
Finally, the classification process of the optimized FNN is analyzed to extract the image features that the FNN uses to decide whether the drop in an image is splashing or nonsplashing.

\subsection{\label{sec:drop_imp_exp}Data collection: drop impact experiment}

A drop impact experiment was carried out to collect high-speed videos of splashing and nonsplashing drops from which images were extracted for image classification.

\subsubsection{\label{sec:exp_setup}Experimental setup}

\begin{figure}[!t]
\centering
\includegraphics[width=\columnwidth]{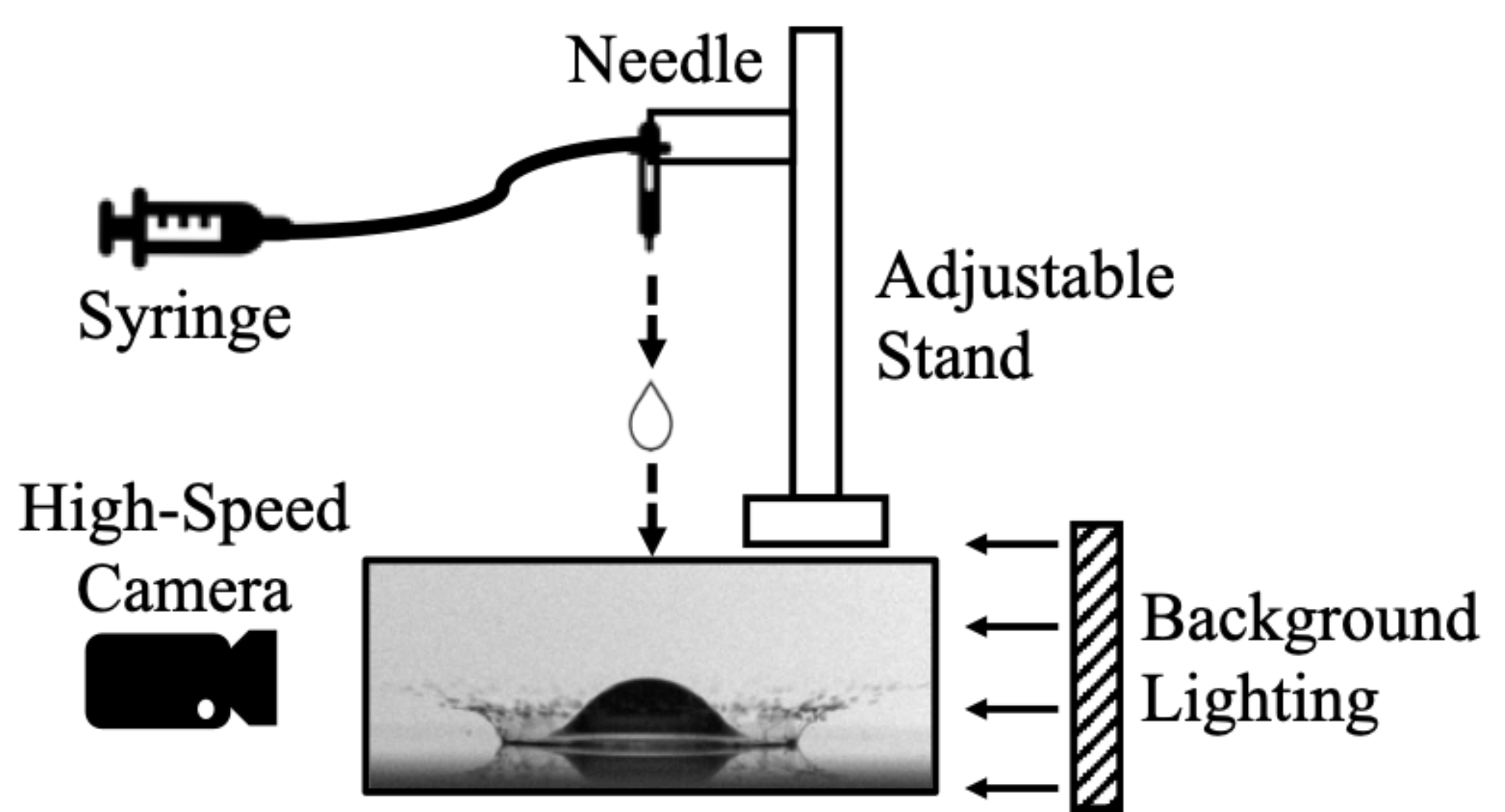}
\caption{\label{fig:setup} Schematic of experimental setup.}
\end{figure}

The experimental setup, shown in Fig.~\ref{fig:setup}, consisted of a syringe, a rubber tube, a plastic needle, an adjustable stand, a glass substrate, a high-speed camera, and background lighting.
The syringe supplied liquid via the rubber tube to the plastic needle (internal diameter 0.97~mm), which was clamped to the adjustable stand. A 
drop formed at the tip of the needle and fell freely before impacting the hydrophilic surface of the glass substrate.
The impact was recorded using the high-speed camera in the presence of background lighting.

%===============================================
%===============================================

\subsubsection{\label{sec:exp_cond}Experimental conditions}

Drop inertia or impact velocity $U_0$ is the only physical property that was manipulated in the experiment.
Through the adjustable stand, $U_0$ was varied by adjusting the impact height $H$, i.e., the vertical distance between the point where the drop started to free fall and the surface of the glass substrate.
$H$ ranged between 4 and 60~cm, and the outcome of the drop impact was either splashing or nonsplashing.

Throughout the experiment, other physical properties of the liquid were kept constant by using the same liquid, ethanol (Hayashi Pure Chemical Ind., Ltd.; density $\rho = 789~\rm{kg/m^3}$, surface tension $\gamma = 2.2 \times 10^{-2}$~N/m, and dynamic viscosity $\mu = 10^{-3}$~Pa$\cdot$s).
The drop size [area-equivalent radius $R_{0} = (1.29 \pm 0.04)\times 10^{-3}$~m] was kept constant by using the same plastic needle.
The physical properties of the solid surface and the ambient air were kept constant by using the same type of glass substrate (Muto Pure Chemicals Co., Ltd., star frost slide glass 511611) and by carrying out the experiment under atmospheric pressure at room temperature ($21$--$25$~$^{\circ}\rm{C}$).

%===============================================
\begin{figure*}[!t]
\subfloat{
\includegraphics[width=0.24\textwidth]{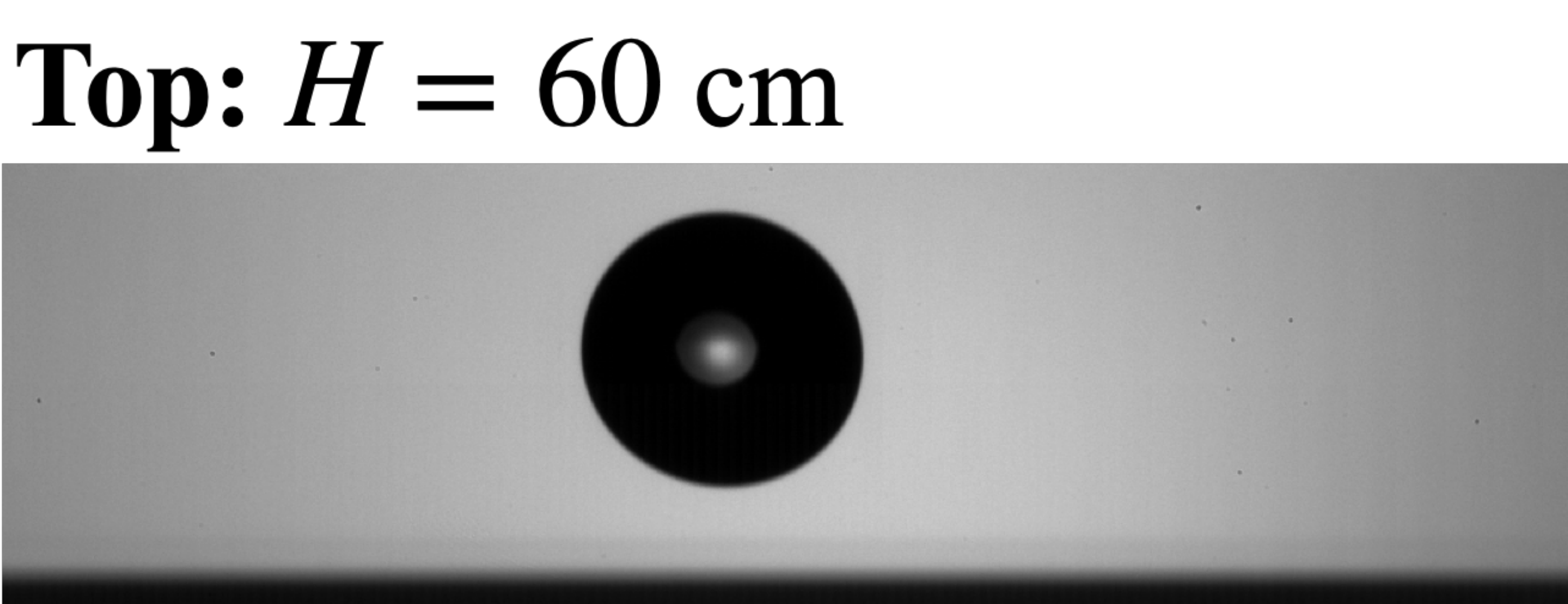}
}
\subfloat{
\includegraphics[width=0.24\textwidth]{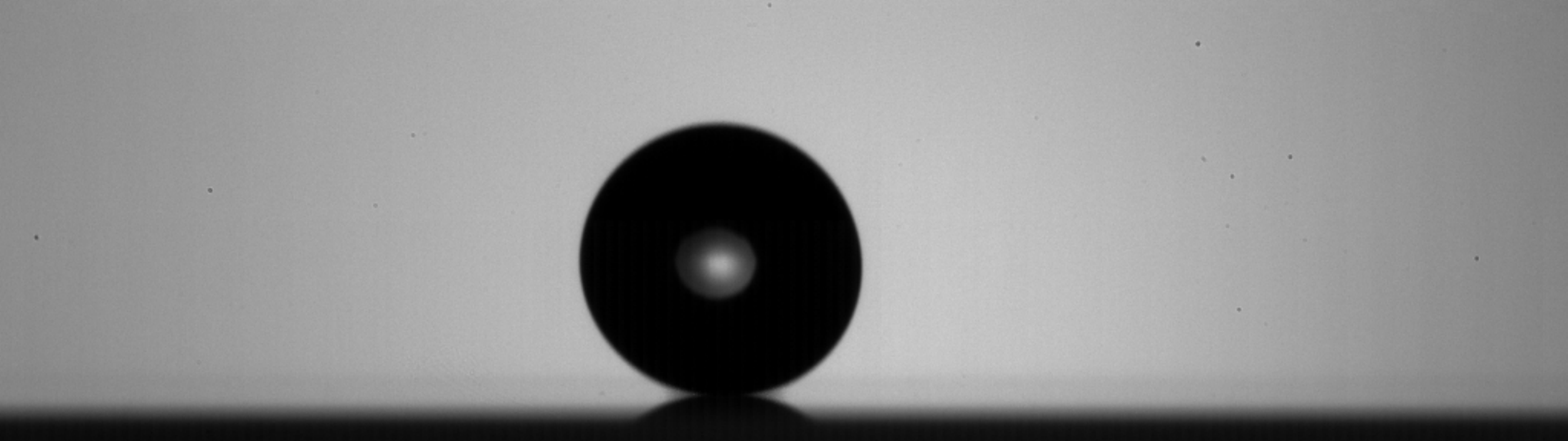}
}
\subfloat{
\includegraphics[width=0.24\textwidth]{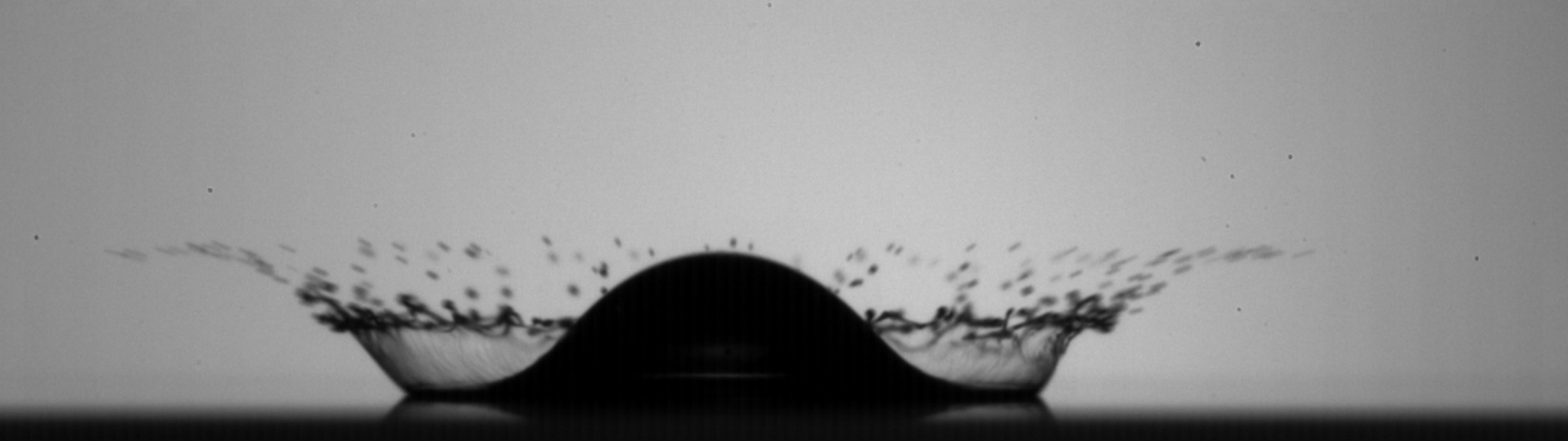}
}
\subfloat{
\includegraphics[width=0.24\textwidth]{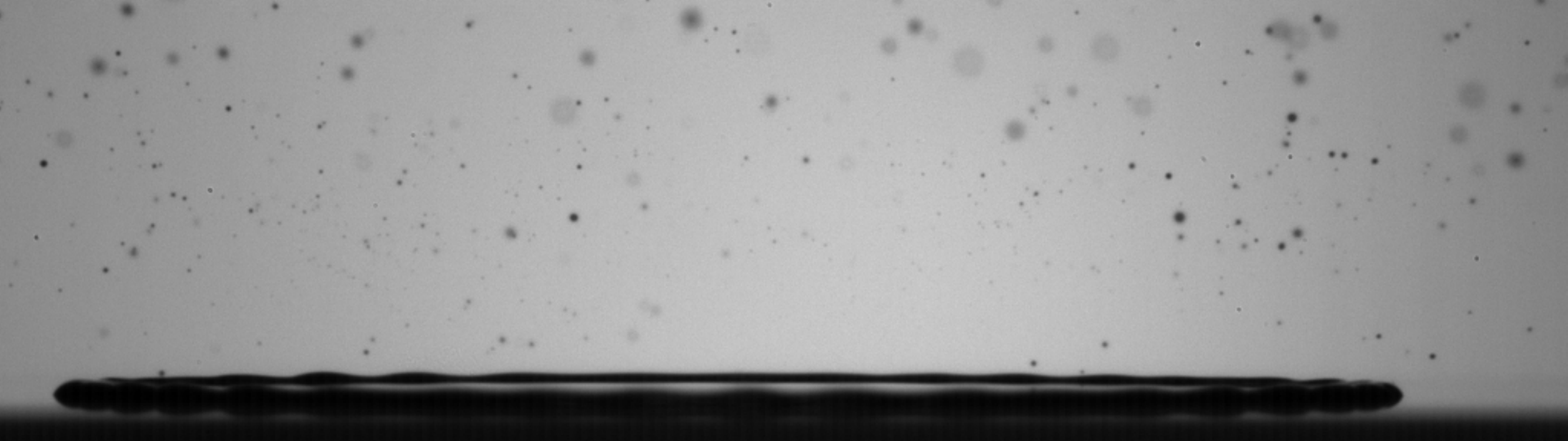}
}\\
\subfloat{
\includegraphics[width=0.24\textwidth]{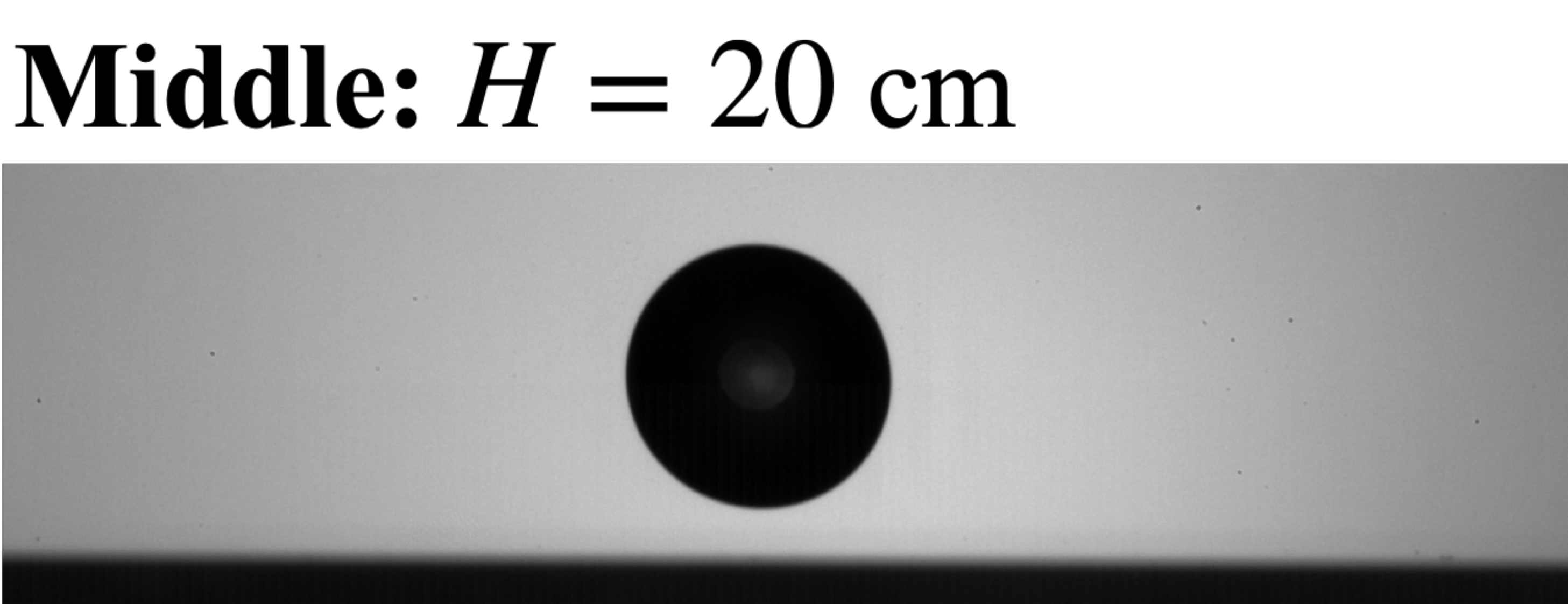}
}
\subfloat{
\includegraphics[width=0.24\textwidth]{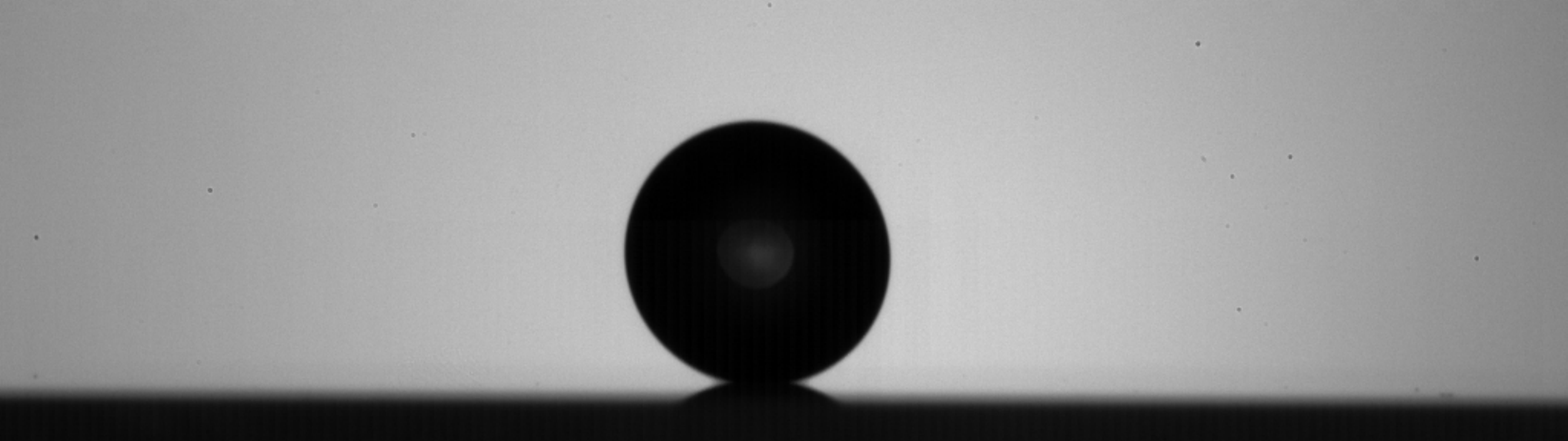}
}
\subfloat{
\includegraphics[width=0.24\textwidth]{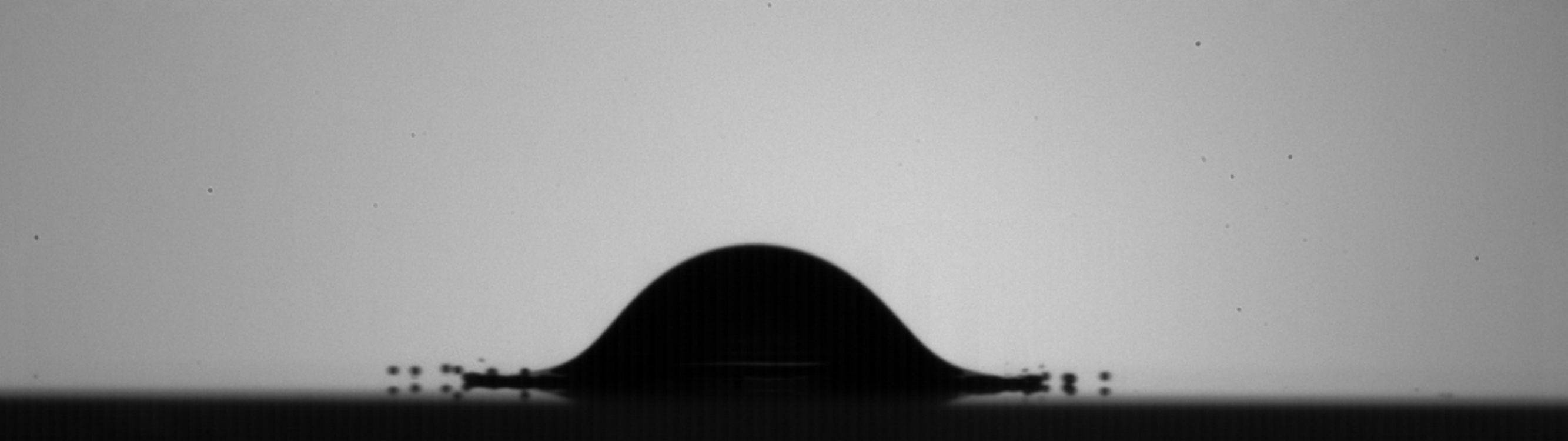}
}
\subfloat{
\includegraphics[width=0.24\textwidth]{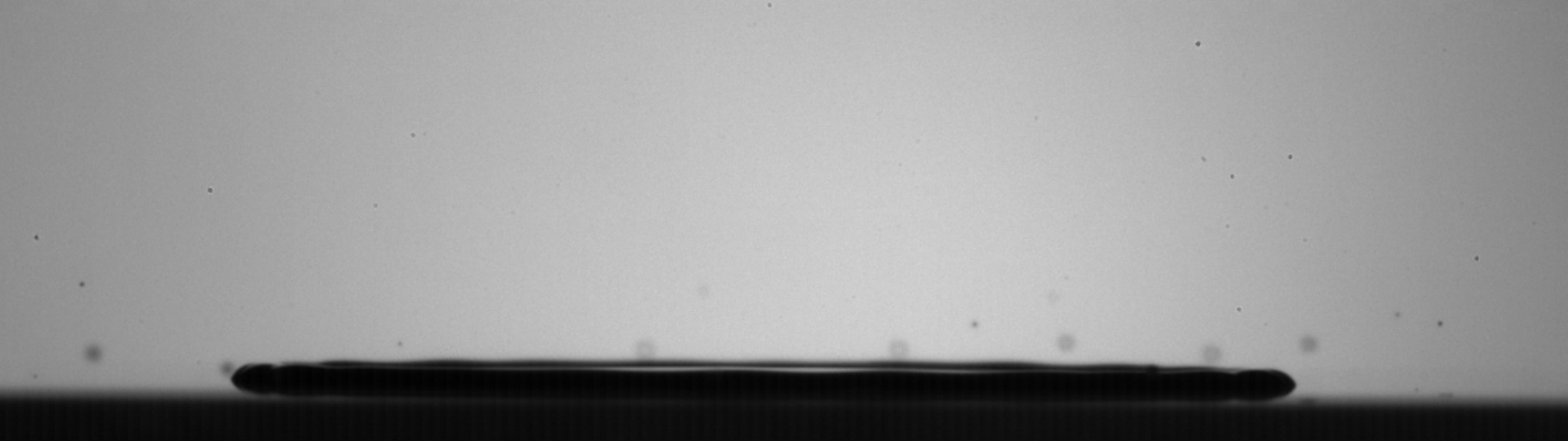}
}\\
\setcounter{subfigure}{0}%This is necessary since otherwise the labeling of the figure parts will be (i), (j), (k), (l)
\subfloat[]{
\includegraphics[width=0.24\textwidth]{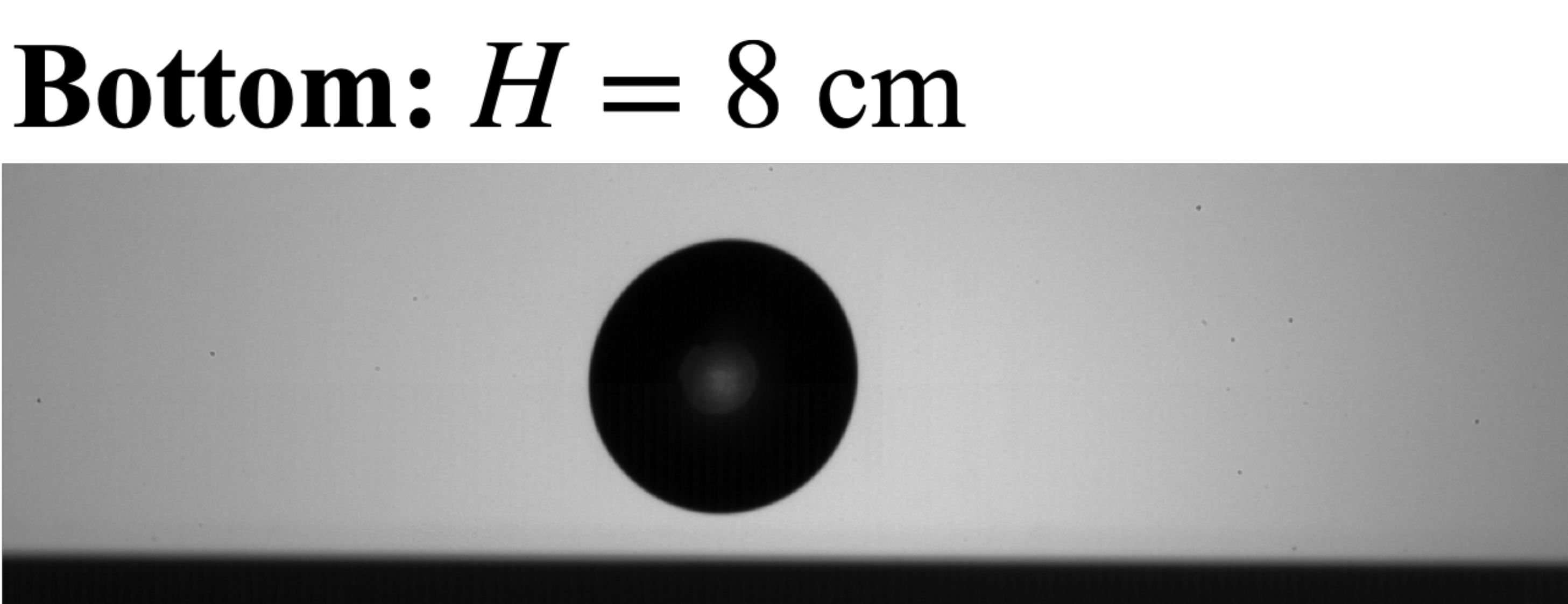}\label{fig:first}
}
\subfloat[]{
\includegraphics[width=0.24\textwidth]{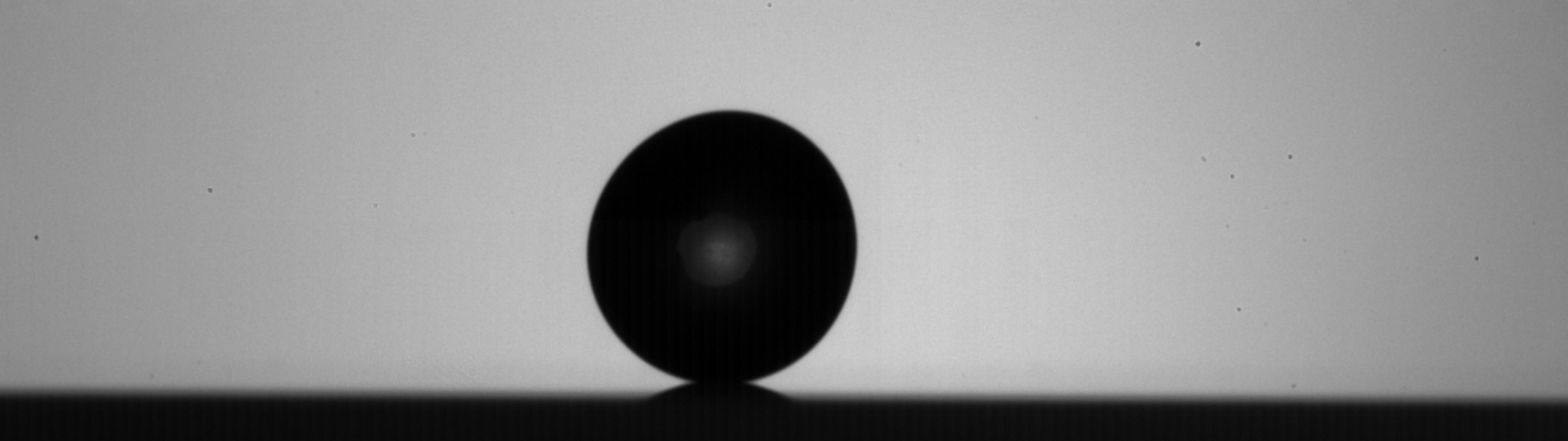}\label{fig:touch}
}
\subfloat[]{
\includegraphics[width=0.24\textwidth]{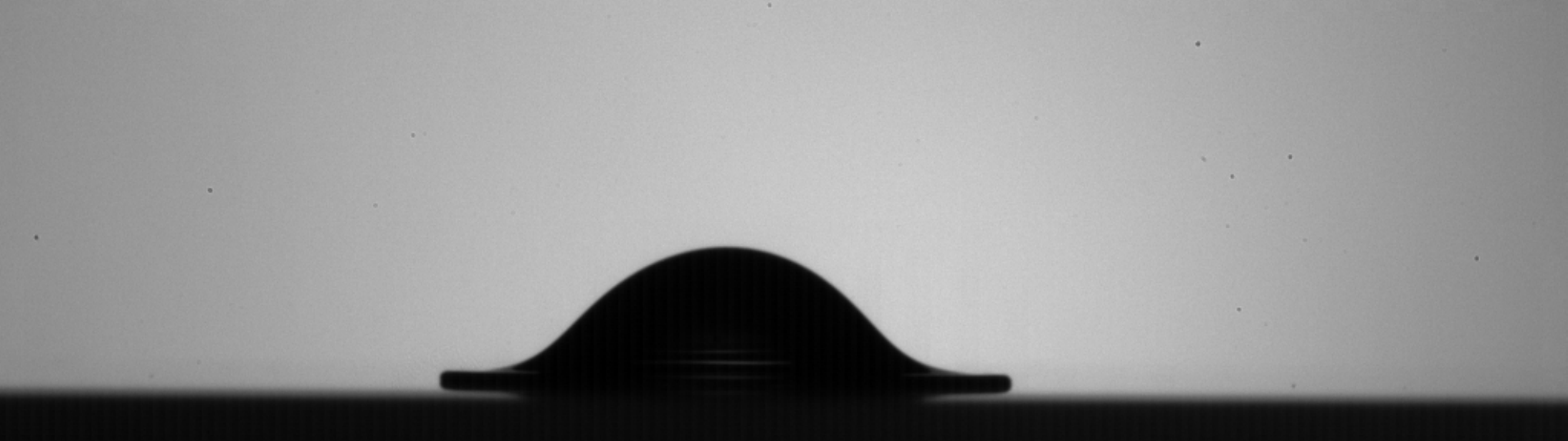}\label{fig:cf}
}
\subfloat[]{
\includegraphics[width=0.24\textwidth]{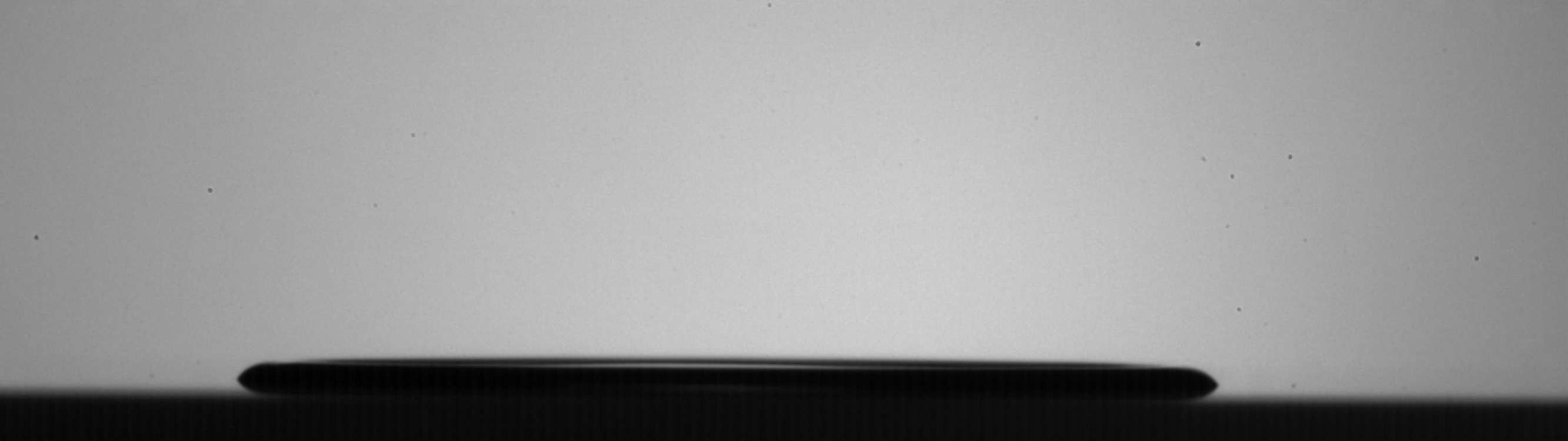}\label{fig:end}
}
\caption{\label{fig:snapshots} Snapshots of the recorded videos for drop impacts with impact height $H = 60$~cm (top), $20$~cm (middle), and $8$~cm (bottom) at different impact times: 
\protect\subref*{fig:first} first recorded frame, which is about $2 \times 10^{-4}$~s before the drop touched the surface; 
\protect\subref*{fig:touch} when the drop first touched the surface; 
\protect\subref*{fig:cf} when half of the drop impacted the surface (which was extracted for image classification using the FNN); 
\protect\subref*{fig:end} when the drop collapsed into a thin sheet of uniform thickness.
}
\end{figure*}
%===============================================

\subsubsection{\label{sec:high_speed_video}High-speed videography}

The drop impact was recorded using a high-speed camera (Photron, FASTCAM SA-X) at 45\,000~frames/s and a spatial resolution of $(1.46 \pm 0.02 )\times 10^{-5}$~m/pixel, in the presence of background lighting.
The recorded videos are sequences of 8-bit grayscale images with an image height of 288~pixels and an image width of 1024~pixels.
The recording started at least nine frames or $2 \times 10^{-4}$~s before the drop touched the surface.
This was to ensure that at the beginning of the recording, the drop did not touch the surface, so that drop size $R_{0}$ and impact velocity $U_0$ could be computed correctly.
The recording ended after the drop deformed into a thin sheet with uniform thickness.
Figure~\ref{fig:snapshots} shows several snapshots of the drop impacts for $H = 60$, 20, and 8~cm.

In the presence of background lighting, during a drop impact, the intensity value is near to 0 at a pixel position that captured the drop or the ejected secondary droplets, since the light is blocked.
On the other hand, the intensity value is approximately equal to the intensity of the background lighting at a pixel position that captured neither the drop nor the ejected secondary droplets.
Note that the intensity value is the luminous intensity captured by the high-speed camera, which scales between 0 and 255 for grayscale images.
To ensure high similarity of the image data, the intensity of the background lighting was set to about 210 for every recording.

A total of 252 videos were recorded: 142 splashing and 110 nonsplashing.

\subsubsection{\label{sec:col_data}Collected data}

\begin{figure}[!b]
\centering
\includegraphics[width=\columnwidth]{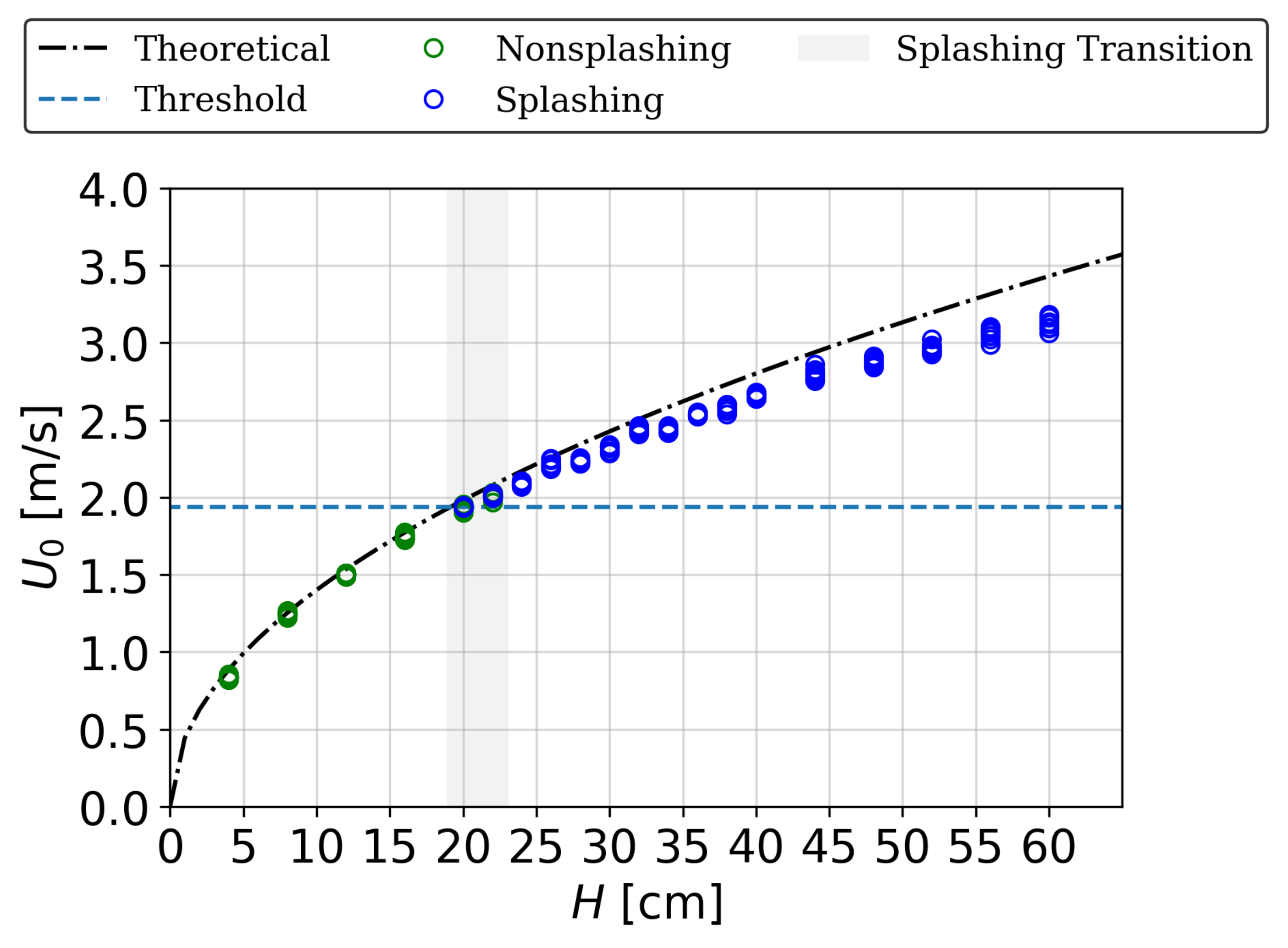}
\caption{\label{fig:U_vs_H}
Impact velocity $U_0$ vs impact height $H$.}
\end{figure}

The outcomes of the impact of ethanol drops that were identified by looking for the presence of secondary droplets and the measured $U_0$ (measured from the ninth frame before impact) after falling from heights $H$ ranging from 4 to 60~cm are summarized in Fig.~\ref{fig:U_vs_H}.
The blue and green circles represent splashing and nonsplashing drop impacts, respectively.
From a frame-by-frame inspection for the presence of secondary droplets by human eyes, it was found that splashing never occurred for drops falling from $H\leq 16$~cm, whereas it always occurred for those falling from $H\geq24$~cm.
In addition, there was a splashing transition at $H = 20$ and 22~cm, where splashing only occurred in a few cases.
As shown in Fig.~\ref{fig:snapshots}, for the splashing that occurred at $H = 20$~cm, only a few secondary droplets were ejected.

%===============================================
%===============================================

The validity of the experimental results was confirmed by comparison with the theoretical drop velocity $U_{0,\mathrm{theo}}$ and the splashing threshold proposed by Usawa \emph{et al.} \cite{usawa2021large} 
The equation for $U_{0,\mathrm{theo}}$ is as follows: 
%===============================================
\begin{equation}
U_{0,\mathrm{theo}} = \sqrt{2gH}
\label{eq:U_0},
\end{equation}
%===============================================
where the gravitational acceleration $g = 9.81~\rm{m/s^2}$.
The curve of $U_{0,\mathrm{theo}}$ against $H$ (black dash-dotted line) shows good agreement with the measured $U_0$.
Note that the measured $U_{0}$ is slightly lower than $U_{0,\mathrm{theo}}$ due to the drag force that acted on the drop during the free fall.
As for the splashing threshold proposed by Usawa \emph{et al.} \cite{usawa2021large}, it is shown in the following equation:
%===============================================
\begin{equation}
K_{l}\mu_{g}U_{t}/\gamma = 0.034
\label{eq:splash_thres},
\end{equation}
%===============================================
where $K_{l}$ is a constant that is deduced using lubrication theory, $\mu_{g}$ is the dynamic viscosity of the surrounding air, and $U_{t}$ is the velocity at which the lamella is initially ejected.
The threshold velocity according to the splashing threshold proposed by Usawa \emph{et al.} \cite{usawa2021large} (horizontal blue dashed line) also validates the outcome of the experiment, since it intersects with the curve of $U_{0,\mathrm{theo}}$ within the experimentally determined splashing transition height.
Such agreements with theory and a previous study prove the validity of the experiment carried out in this study.
In the remaining sections of this article, the drop inertia is presented in the form of the impact height $H$.

%===============================================
\begin{table}[!t]
\caption{\label{tab:dim_num} Ranges of dimensionless numbers.}
\centering
\begin{tabular}{lc}
\hline\hline
Froude number $Fr$ & $7$--$31$ \\ 
Ohnesorge number $Oh$ & $6\times10^{-3}$--$7\times10^{-3}$ \\ 
Reynolds number $Re$ & $7.7\times10^{2}$--$3.0\times10^{3}$ \\ 
Stokes number $St$ & $4.6\times10^{4}$--$1.8\times10^{5}$ \\ 
Weber number $We$ & 31--472\\
\hline\hline
\end{tabular}
\end{table}
%===============================================

The corresponding dimensionless numbers: Froude number $Fr$ $(= U_{0}/\sqrt{gR_{0}})$, Ohnesorge number $Oh$ $(= \mu/\sqrt{\rho\sigma R_{0}})$, Reynolds number $Re$ $(= \rho U_{0} R_{0}/\mu)$, Stokes number $St$ $(= \rho U_{0} R_{0}/\mu_{g})$, and Weber number $We$ $(= \rho U_{0}^2 R_{0}/\gamma)$ are shown in Table~\ref{tab:dim_num}.

\subsection{\label{sec:data_prep}Data preparation}

\subsubsection{\label{sec:dip}Digital image processing}

%===============================================
\begin{figure*}[!t]
\centering
\includegraphics[width=0.8\textwidth]{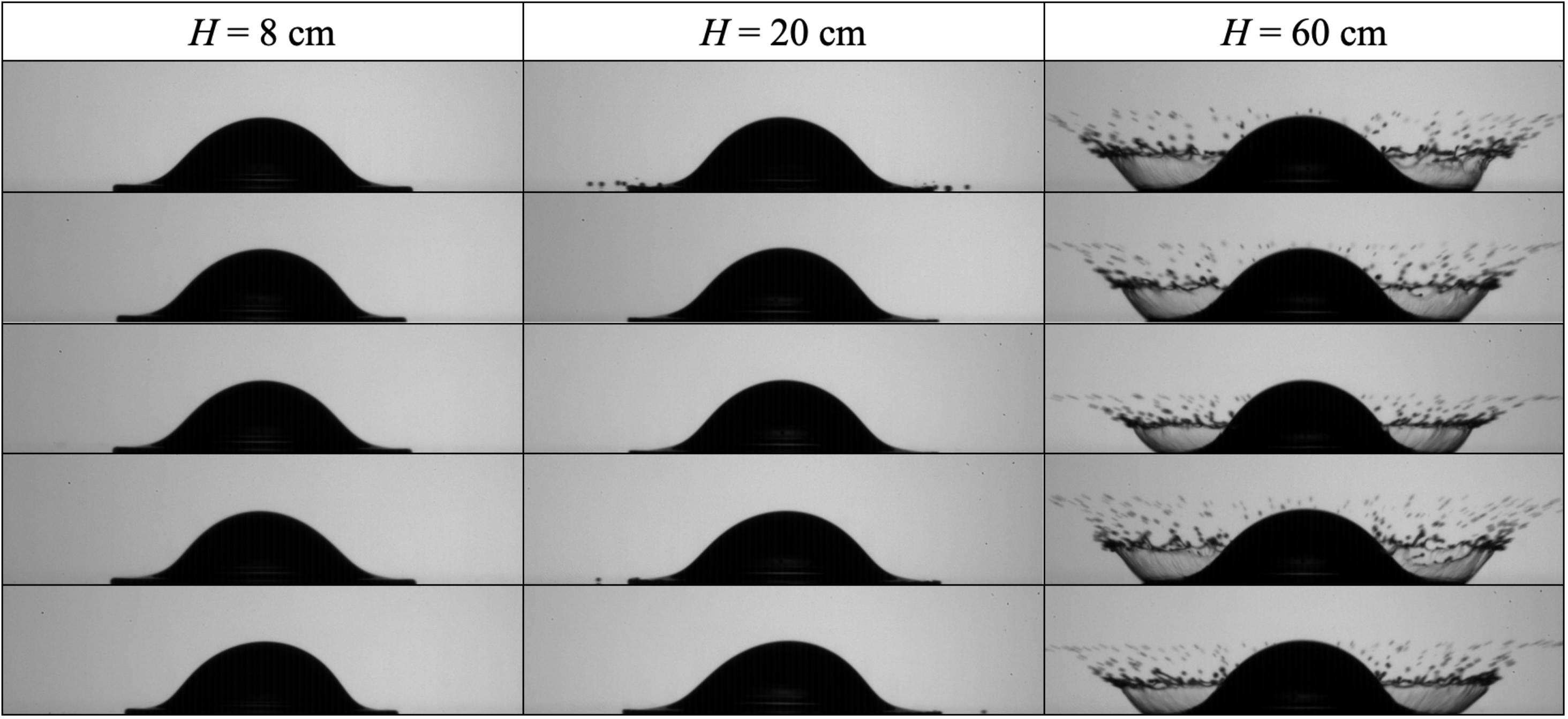}
\caption{\label{fig:processed_img} 
Several examples of the processed images for the drop impacts with impact height $H = 8$, 20, and 60~cm.
The images in the first row are the processed images of the images shown in Figure~\ref{fig:cf}.
}
\end{figure*}
%===============================================

Digital image processing was performed to prepare highly similar and high-quality images for image classification using an FNN.
This included frame extraction and removal of the image background.

In this study, the frame in which half of the drop impacted the surface, i.e., when the central height of the drop normalized by the area-equivalent diameter was $z_0/2R_{0} = 0.5$, was selected from each recorded drop impact video to train, validate, and test the FNN.
The frame when $z_0/2R_{0} = 0.5$ was extracted using an in-house MATLAB code.
This MATLAB code executed image processing that included binarization \cite{otsu1979threshold}, nonlocal means filtering \cite{buades2005non}, and object analyses such as circle detection \cite{yuen1990comparative, atherton1999size} and edge detection \cite{canny1986computational}.
Examples of the extracted images are shown in Fig.~\ref{fig:cf}.

The same code also cropped the extracted images from $288 \times 1024$~pixel$^2$ to $160 \times 640$~pixel$^2$, with the impacting drop and the substrate surface at the center and bottom of the cropped image, respectively.
This is to reduce the computation time and to increase the interpretability of the image classification process.

As a result of similar videographing conditions and the digital image processing explained in this subsection, the images are highly similar in terms of drop height and background lighting, regardless of $H$ and the outcomes (splashing/nonsplashing).
Figure~\ref{fig:processed_img} shows several examples of the processed images for the drop impacts with $H = 60$, 20, and 8~cm.
The images in the first row are the processed images of the images shown in Figure~\ref{fig:cf}.

\subsubsection{\label{sec:data_segment}Data segmentation for cross-validation}

%===============================================
\begin{table*}[!ht]
\caption{
\label{tab:data_num} Respective numbers of splashing and nonsplashing data for training--validation and testing of each data combination.
}
\centering
\begin{tabular}{cccccccc}
\hline\hline
\multirow{3}{*}{Combination}
& \multicolumn{7}{c}{Number of data} \\
\cline{2-8}
& \multicolumn{3}{c}{Training--validation} & \multicolumn{3}{c}{Testing} & \multirow{2}{*}{Total} \\
\cline{2-4}\cline{5-7}
& Splashing& Nonsplashing& Total& Splashing& Nonsplashing& Total& \\\hline
1& 114 & 87& 201& 27& 21& 48& 249\\ 
2& 112& 86& 198& 29& 22& 51& 249\\ 
3& 113& 85& 198& 28& 23& 51& 249\\ 
4& 114& 85& 199& 27& 23& 50& 249\\ 
5& 111& 89& 200& 30& 19& 49& 249\\
\hline\hline
\end{tabular}
\end{table*}
%===============================================

To ensure the quality of the trained FNN, cross-validation, which includes training, validation, and testing of the FNN, was implemented \cite{cawley2010over,geron2019hands}.
The purpose of training was to find the best set of weights and biases for the trained model.
Validation was carried out concurrently with training for hyperparameter tuning and prevention of underfitting and overfitting.
Last, but not least, testing was done to ensure the generalizability of the trained FNN, i.e., its ability to classify new images that were not used for training.

For implementation of cross-validation, the collected data (labeled images) were segmented into training, validation, and test data.
For fivefold cross-validation, the collected data were first divided into five different groups, with each group consisting of about 20\% of the collected data.
One of these groups was kept aside and reserved for testing, while the remaining roughly 80\% of the data were used to train and validate the FNN.
Out of this 80\% training--validation data, about 10\% were picked randomly for validation, and thus only about 70\% were used for training.
Consequently, there were five different combinations of training--validation--test data that were used to train--validate and test the FNN.
Thus, there were five different sets of results to be analyzed.
To ensure that the data for every impact height $H$ were included in both the training--validation and the test datasets for all five data combinations, 80\%--20\% segmentation was carried out for each $H$ before they were pooled into the respective data combinations.
As shown in Table~\ref{tab:data_num}, the numbers of splashing and nonsplashing data for training--validation and testing are similar regardless of the data combination, i.e, about 200 and 50 for training--validation and testing, respectively.
To check whether the number of data is sufficient, training was performed using a smaller number of training--validation data.
The results are similar even when the segmentation is 20\% for training--validation and 80\% for testing, thus confirming that the number of data is sufficient for the objectives of this study.

%===============================================
\begin{figure*}[!ht]
\centering
\includegraphics[width=0.64\textwidth]{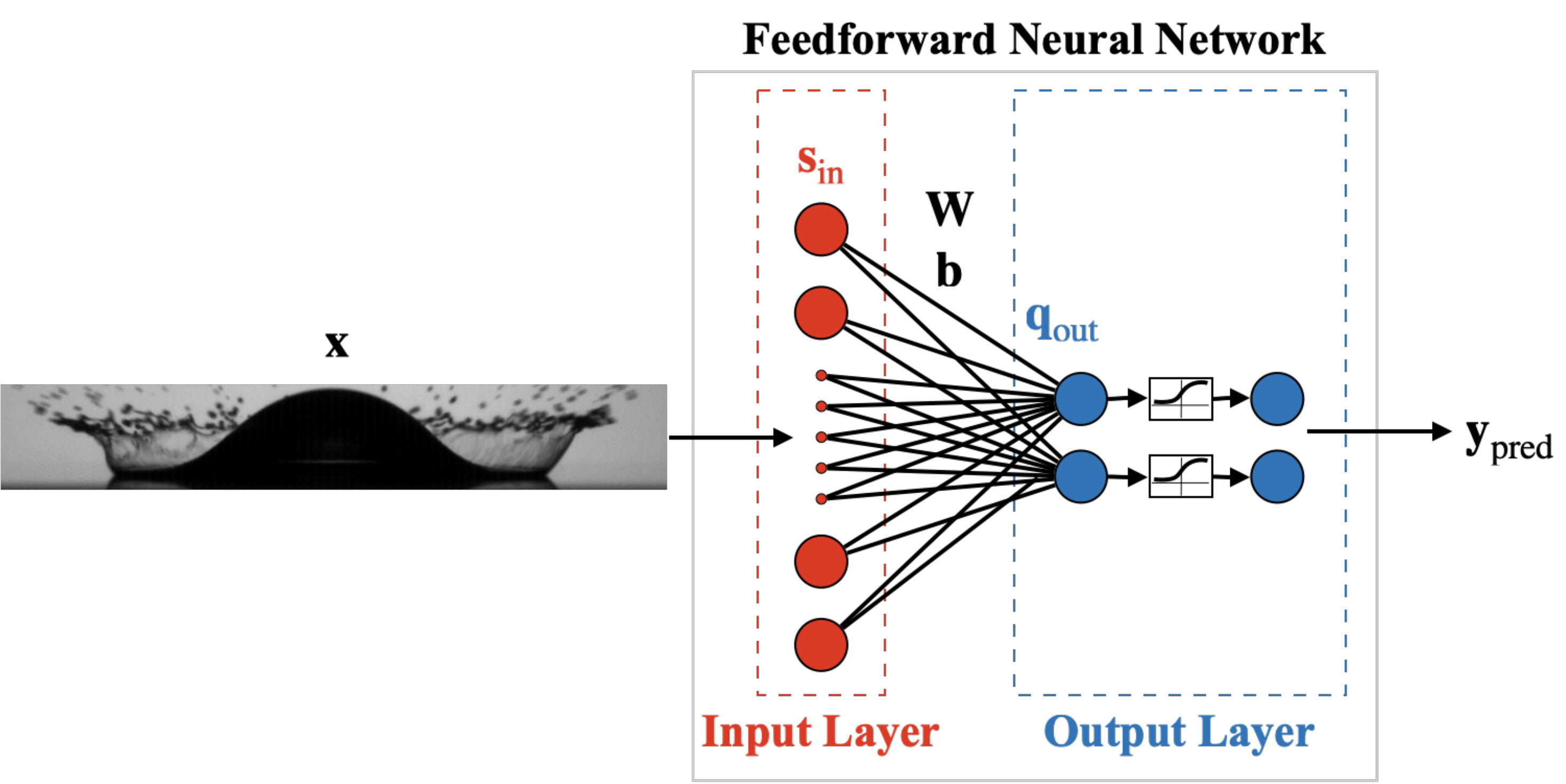}
\caption{\label{fig:architecture} Optimized feedforward neural network (FNN) architecture for classification of splashing and nonsplashing images.}
\end{figure*}
%===============================================
%===============================================
\begin{figure*}[!ht]
\centering
\includegraphics[width=0.8\textwidth]{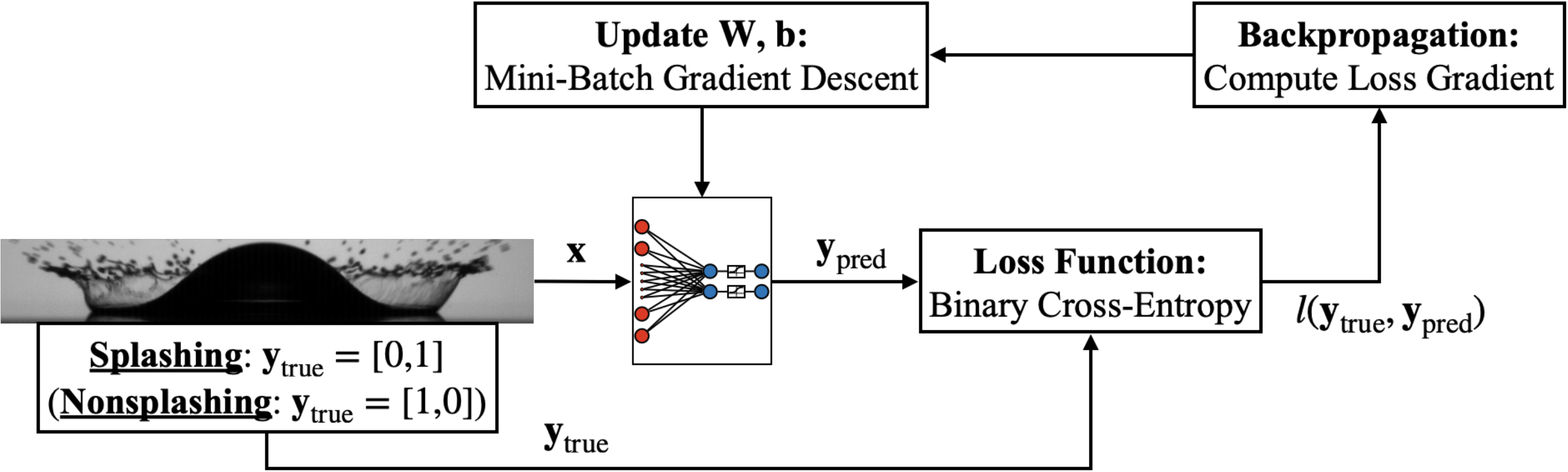}
\caption{\label{fig:train} Flowchart of feeforward neural network (FNN) training.}
\end{figure*}
%===============================================

\subsection{\label{sec:fnn}Image classification: feedforward neural network (FNN)}

In this subsection, the details of the FNN that was used to classify the images of splashing and nonsplashing drops are explained.
The implementation of the FNN was done in the Python programming language on Google Colaboratory \cite{carneiro2018performance} using the libraries of TensorFlow \cite{abadi2016tensorflow}.
Through architecture optimization (the process of which is explained in Appendix~\ref{app:arch_opt}), an FNN with zero hidden layer was chosen.
In Sec.~\ref{sec:fnn_arch}, the details of the optimized architecture and the mathematical operations involved in the FNN are given, and 
in Sec.~\ref{sec:train_val_test}, the algorithms and mathematical equations involved in the training, validation, and testing of the FNN are described.

\subsubsection{\label{sec:fnn_arch}Neural network architecture}

The optimized architecture of the FNN is shown in Fig.~\ref{fig:architecture}.
The optimized FNN with no hidden layer exhibited a classification performance as high as that of an FNN with hidden layers while being superior to the later in terms of higher interpretability and lower computational cost.
Since there is no hidden layer, the input layer is fully connected to the output layer.

As mentioned in Sec.~\ref{sec:dip}, the input images were images cropped into size $h_\mathrm{img}\times w_\mathrm{img} = 160\times 640$~pixels$^2$.
In the input layer, these were flattened in row-major order from two-dimensional matrices and transposed into one-dimensional column vectors: $\mathbf{x} \in \mathbb{R}^{h_\mathrm{img}\times w_\mathrm{img}} \to \mathbf{s}_\mathrm{in} \in \mathbb{R}^{M}$, for $M = h_\mathrm{img}\times w_\mathrm{img}$.
The value of each element in the vectors was normalized from 0--255 to 0--1.

Each element of ${\bf s}_\mathrm{in}$ (red circles in Fig.~\ref{fig:architecture}) is connected to each element of ${\bf q}_\mathrm{out}$ in the output layer (blue circles) by a linear function:
%===============================================
\begin{equation}
{\bf q}_\mathrm{out} = {\bf W}{\bf s}_\mathrm{in} + {\bf b}
\label{eq:lin_func_out},
\end{equation}
%===============================================
where ${\bf q}_\mathrm{out} \in \mathbb{R}^{C}$ is the result of this mathematical operation, ${\bf W} \in \mathbb{R}^{C \times M}$ is the weight that connects the input and output layers, and ${\bf b} \in \mathbb{R}^{C}$ is the bias vector.
$C$ is the number of output classes, which are splashing and nonsplashing in this case, and so $C=2$.
Both ${\bf W}$ and ${\bf b}$ are updated through neural network training.

Each element of ${\bf q}_\mathrm{out}$ was set to be activated by a sigmoid function:
%===============================================
\begin{equation}
y_{\mathrm{pred},i} = \sigma(q_{\mathrm{out},i}) = \frac{1}{1+e^{-q_{\mathrm{out},i}}}
\label{eq:sigmoid},
\end{equation}
%===============================================
for $i = 1,\dots,C$, where ${\bf y}_\mathrm{pred} \in \mathbb{R}^{C}$ is the result of this mathematical operation.
As shown in Eq.~(\ref{eq:sigmoid}), a sigmoid function saturates negative values at 0 and positive values at 1.
Thus, ${\bf y}_\mathrm{pred} = [y_{\mathrm{pred},1}, y_{\mathrm{pred},2}]$ can be interpreted as a vector that contains the probabilities of an input image to be classified as a nonsplashing drop $y_{\mathrm{pred},1}$ and as a splashing drop $y_{\mathrm{pred},2}$.
For binary classifications, the sum of $y_{\mathrm{pred},1}$ and $y_{\mathrm{pred},2}$ is approximately equal to 1.

The classification threshold of the trained FNN was set to be 0.5.
In other words, the trained FNN classifies an image based on the element of ${\bf y}_\mathrm{pred}$ that has a value equal to or greater than 0.5.
For example, if the prediction of an image by the trained FNN is ${\bf y}_\mathrm{pred} = [0.20,0.80]$, then the image will be classified as an image of a splashing drop.

\subsubsection{\label{sec:train_val_test}Training and validation}

%===============================================
\begin{figure*}[!ht]
\subfloat[]{
\includegraphics[width=\columnwidth]{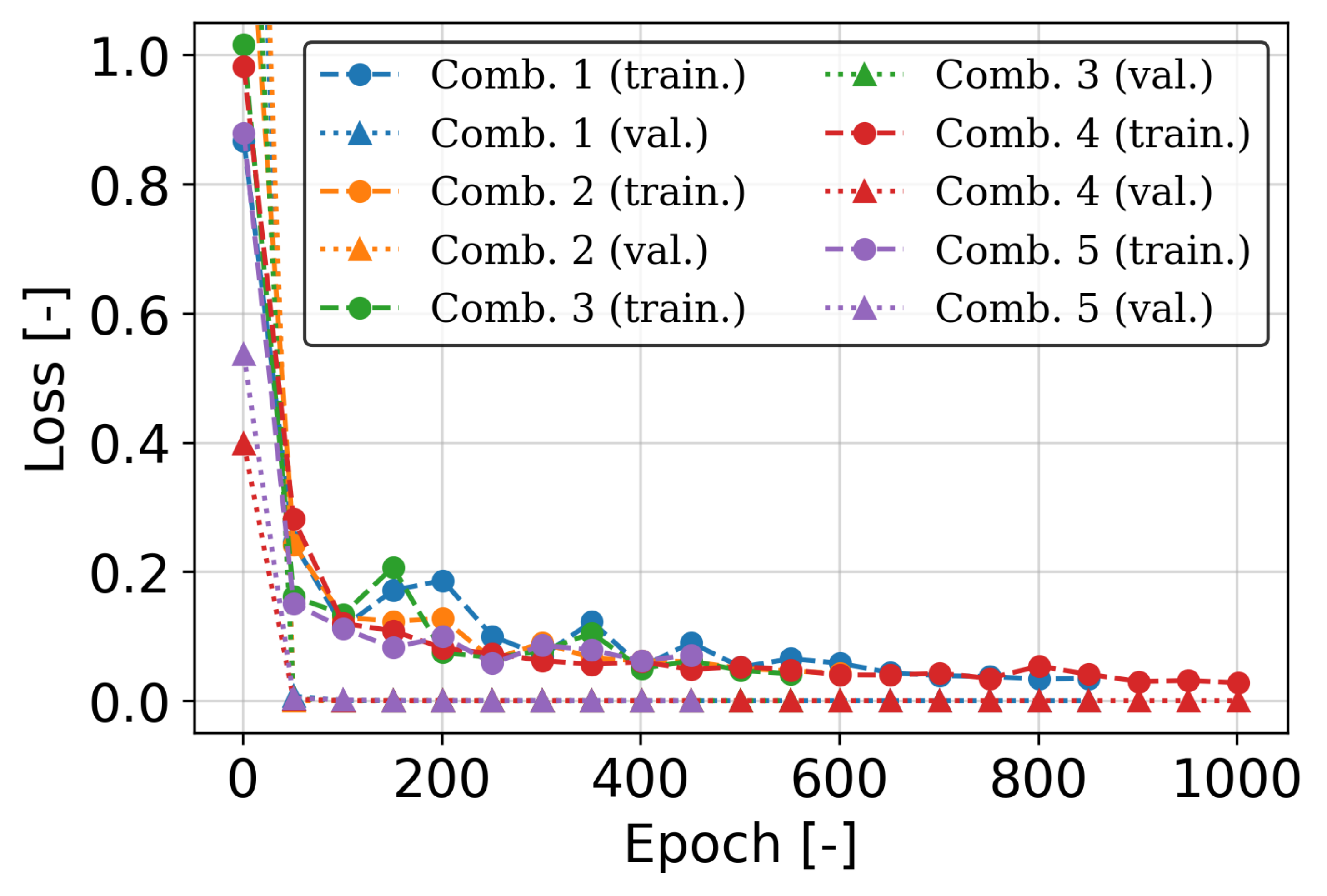}
\label{fig:loss} }
\hfill
\subfloat[]{
\includegraphics[width=\columnwidth]{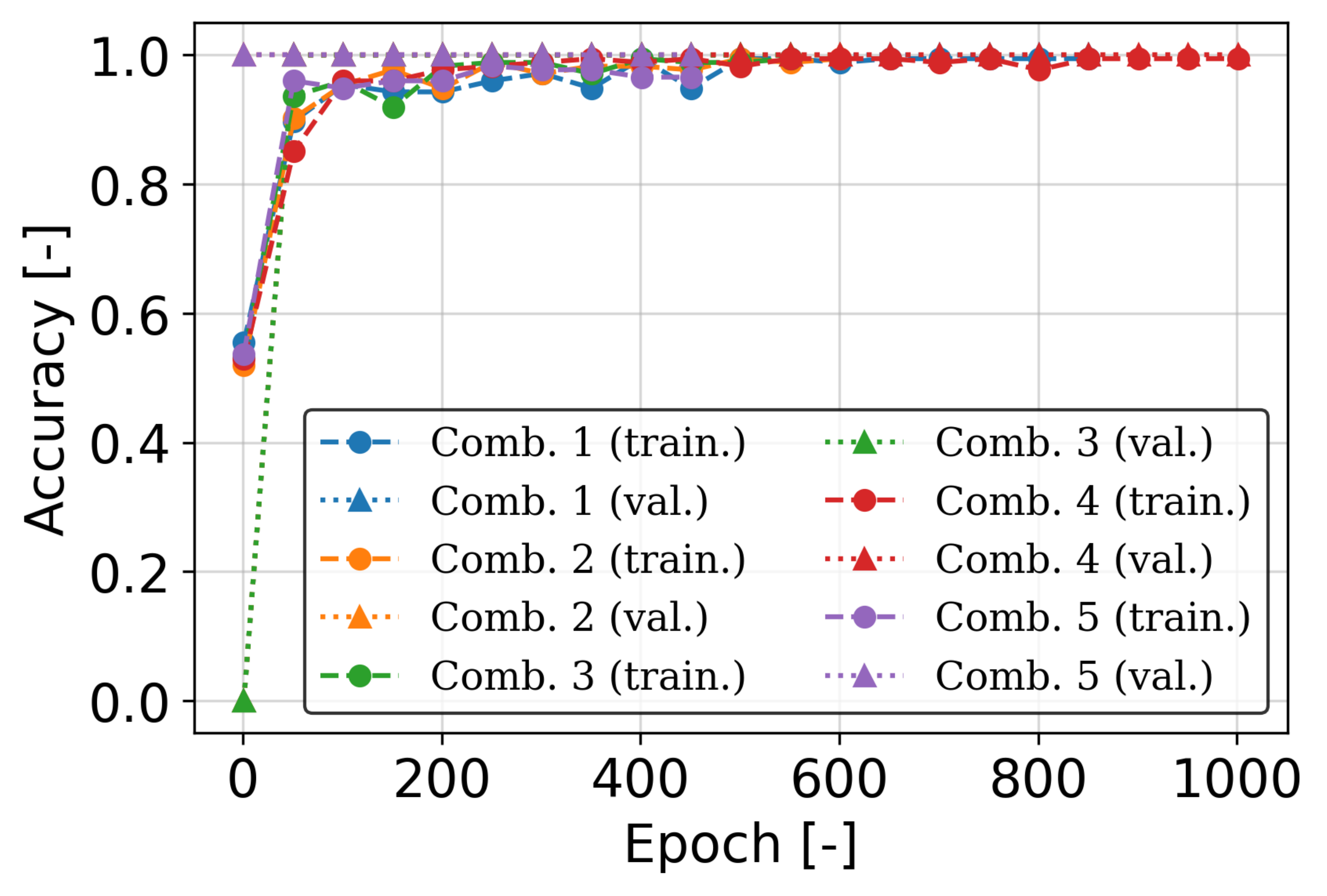}
\label{fig:acc} }
\caption{\label{fig:class_perform} 
Evaluation of the training of the FNNs: \protect\subref*{fig:loss} loss and \protect\subref*{fig:acc} accuracy after every fifty epochs.
Comb., combination; train., training loss or accuracy; val.. validation loss or accuracy).
}
\end{figure*}
%===============================================

The purpose of neural network training is to determine the value of each element in the weight matrix ${\bf W}$ and the bias vector ${\bf b}$ of the FNN, which were initialized using the Glorot uniform initializer \cite{glorot2010understanding}.
The training process is illustrated by the flowchart shown in Fig.~\ref{fig:train}.

Every trained image is fed through the FNN to compute ${\bf y}_\mathrm{pred}$, which is then compared with the label of the trained image ${\bf y}_\mathrm{true}\in \mathbb{R}^{C}$.
As already mentioned in Sec.~\ref{sec:col_data}, the images were inspected frame-by-frame and labeled with ${\bf y}_\mathrm{true}$.
For an image of a splashing drop, ${\bf y}_\mathrm{true} = [0,1]$, while for an image of a nonsplashing drop, ${\bf y}_\mathrm{true} = [1,0]$.

The comparison between ${\bf y}_\mathrm{pred}$ and ${\bf y}_\mathrm{true}$ is made by computing the loss $l$ using the cross-entropy loss function for binary classification, as follows:
%===============================================
\begin{multline}
l({\bf y}_\mathrm{true},{\bf y}_\mathrm{pred})\\
=\sum_{i=1}^{C}\left[
-{y}_{\mathrm{true},i}\ln({ y}_{\mathrm{pred},i})-(1-{y}_{\mathrm{true},i})\ln(1-{y}_{\mathrm{pred},i})\right]
\label{eq:cross_entropy},
\end{multline}
%===============================================
for $i = 1, \dots, C$.
If ${\bf y}_\mathrm{pred}$ is close to ${\bf y}_\mathrm{true}$, then $l$ will be close to 0.
However, if ${\bf y}_\mathrm{pred}$ is not equal to ${\bf y}_\mathrm{true}$, $l$ will increase dramatically as ${\bf y}_\mathrm{pred}$ deviates further from ${\bf y}_\mathrm{true}$.
The loss function was used to evaluate the model both during training and validation, but not during testing.

From this computed loss $l$, a backpropagation algorithm \cite{rumelhart1986learning} was applied to compute the gradient of the loss function with respect to each element of $\bf W$ and $\bf b$ of the FNN.
With the computed gradients, the algorithm can determine how each element of $\bf W$ and $\bf b$ should be tweaked to minimize the loss.
To tweak them in the direction of descending gradients, an algorithm called mini-batch gradient descent \cite{li2014efficient} was used.

As well as the cross-entropy loss function [Eq.~(\ref{eq:cross_entropy})], the classification accuracy of the FNN was also evaluated:
%===============================================
\begin{equation}
\text{accuracy}= \frac{\text{number of correct predictions}}{\text{total number of predictions}}.
 \end{equation}
%===============================================
The number of correct predictions is determined by the classification threshold, which was set to 0.5.
The accuracy of the model was evaluated during training, validation, and testing.

To avoid overfitting, a regularization technique called early stopping \cite{prechelt1998early} was applied.
The training of the FNN was evaluated from the plots of loss and accuracy against number of epochs, which are shown in Fig.~\ref{fig:class_perform}.
For better visibility, only the loss and accuracy after every fifty epochs are plotted in the figure.
Here, the number of epochs indicates how many times all training images are fed through the FNN for training.
As the number of epochs increased, both training and validation losses decreased and approached 0.
With early stopping, the losses did not increase after reaching their minimum value.
On the other hand, both training and validation accuracies increased with increasing number of epochs and eventually reached 1.
The same trend was observed for all data combinations.
This showed that the training was valid and the trained FNN was well generalized.
The trained FNN was then tested for its accuracy in classifying test images.

\section{\label{sec:result}Results and Discussion}

In this section, the results and an in-depth discussion are presented.
In Sec.~\ref{sec:class_perform}, the testing of the trained FNN is explained.
In Sec.~\ref{sec:visual}, the process for extracting the image features used by the FNN for classification is elaborated.
Finally, in the discussion in Sec.~\ref{sec:class_criteria}, an attempt is made to understand the physical interpretation of the extracted image features.

\subsection{\label{sec:class_perform}Testing of FNN}

%===============================================
\begin{table*}[!ht]
\caption{\label{tab:test_result} Test accuracy of FNN trained with each data combination.}
\centering
\begin{tabular}{crrrrrr}
\hline\hline
\multirow{2}{*}{Combination}& 
\multicolumn{6}{c}{Test accuracy} \\\cline{2-7} &
\multicolumn{2}{c}{Splashing} & \multicolumn{2}{c}{Nonsplashing} & \multicolumn{2}{c}{Total} \\ \hline
1& 27/27& 100\%& 21/21& 100\%& 48/48& 100\% \\
2& 26/29& 90\%& 22/22& 100\%& 48/51& 94\% \\ 
3& 26/28& 100\%& 22/23& 96\%& 48/51& 94\% \\
4& 26/27& 96\%& 23/23& 100\%& 49/50& 98\% \\ 
5& 28/30& 93\%& 19/19& 100\%& 47/49& 96\% \\
\hline\hline
\end{tabular}
\end{table*}
%===============================================

%===============================================
\begin{figure}[!b]
\centering
\includegraphics[width=\columnwidth]{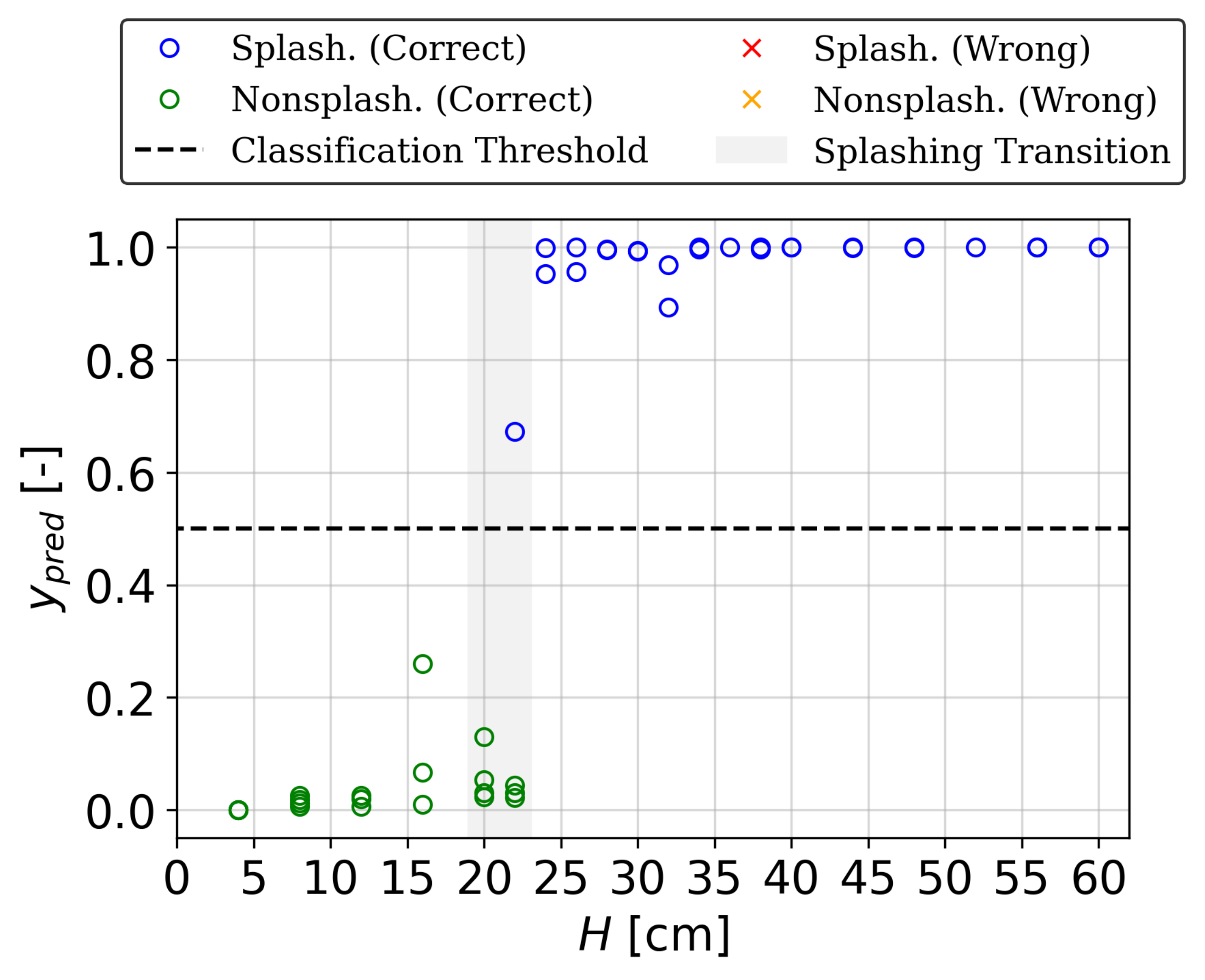}
\caption{\label{fig:spl_prob} Splash probability predicted by the FNN, $y_{\mathrm{pred},2}$, for test images with different impact heights $H$ (combination 1).}
\end{figure}
%===============================================

%===============================================
\begin{figure*}[!ht]
\centering
\includegraphics[width=0.8\textwidth]{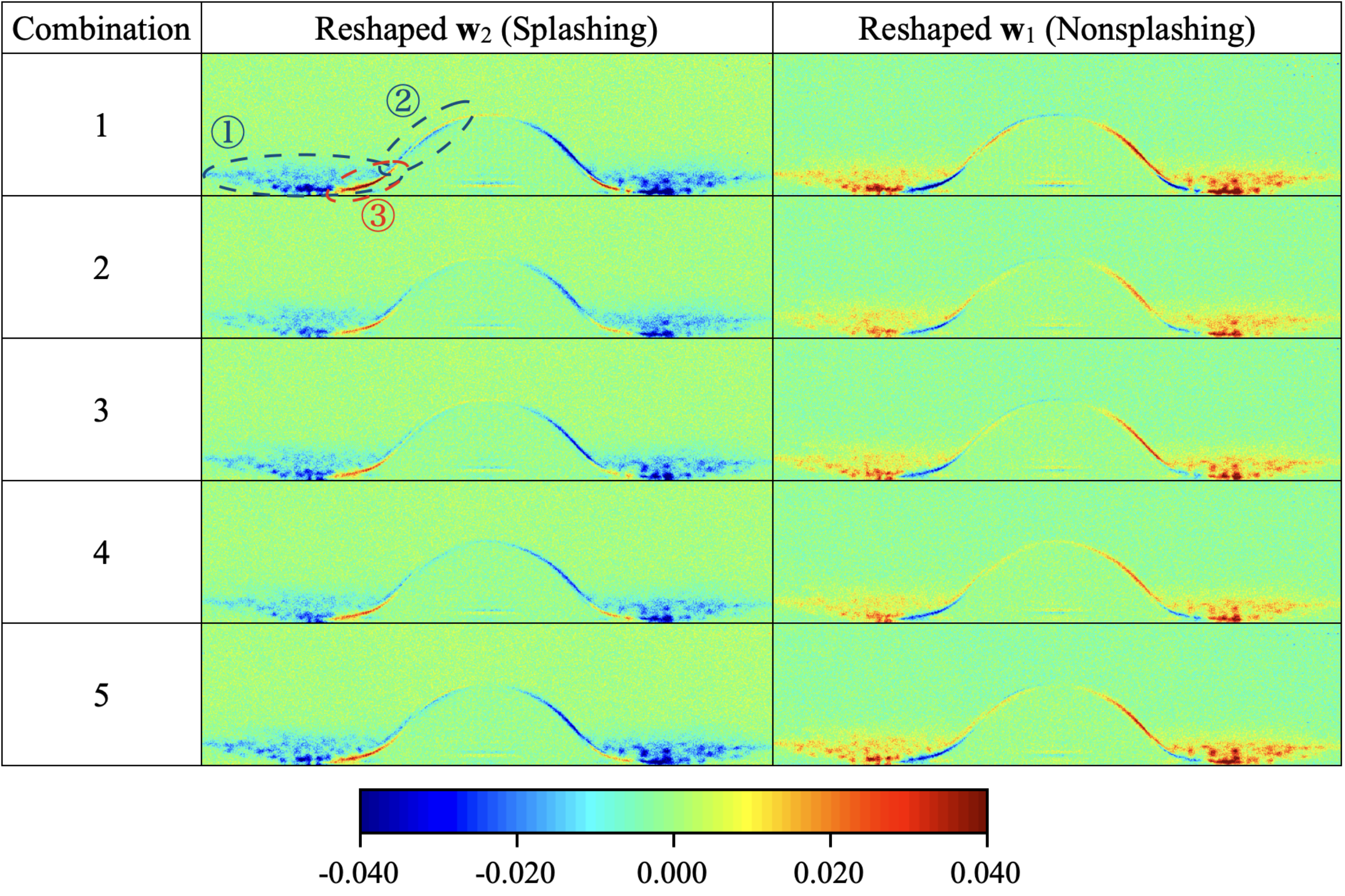}
\caption{
\label{fig:trained_w}
Colormaps of reshaped ${\mathbf w}_{2}$ and ${\mathbf w}_{1}$ of the FNN trained with each data combination.
\textcircled{1}, \textcircled{2}, and \textcircled{3} indicate the distribution of values with large magnitude in ${\mathbf w}_{2}$ of the FNN trained with combination 1, which are symmetric.
Such distribution is same as ${\mathbf w}_{1}$ but with opposite signs.
Despite being trained with different data combinations, ${\mathbf w}_{2}$ and ${\mathbf w}_{1}$ are similar for all FNNs, indicating good generalizability.
}
\end{figure*}
%===============================================
%===============================================
\begin{figure}[!ht]
\subfloat[]{
\includegraphics[width=0.9\columnwidth]{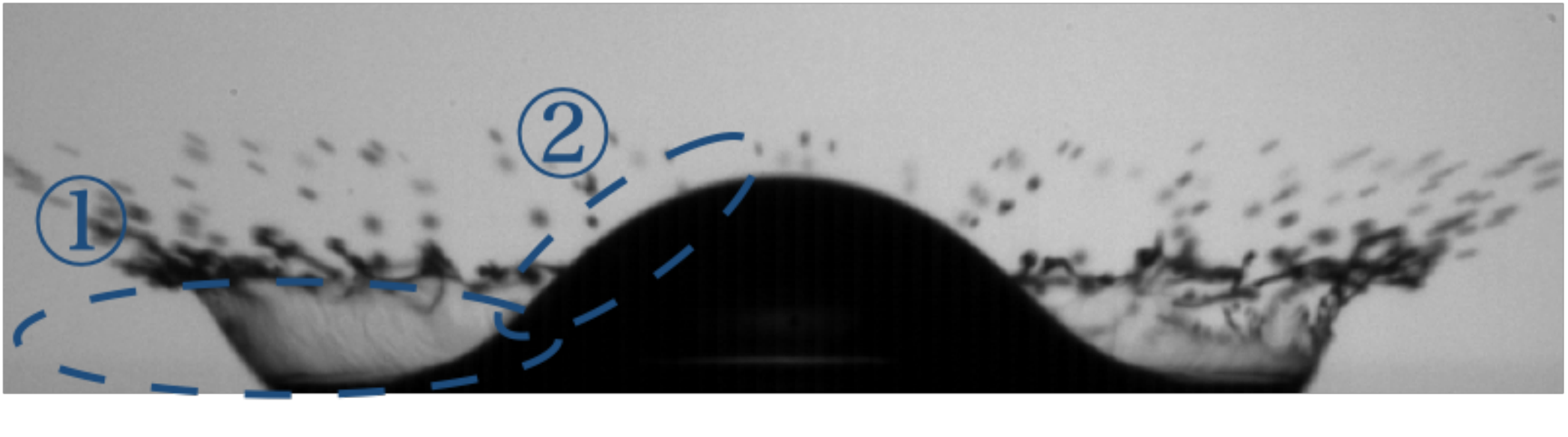}
\label{fig:251_ori} }\\
\subfloat[]{
\includegraphics[width=0.9\columnwidth]{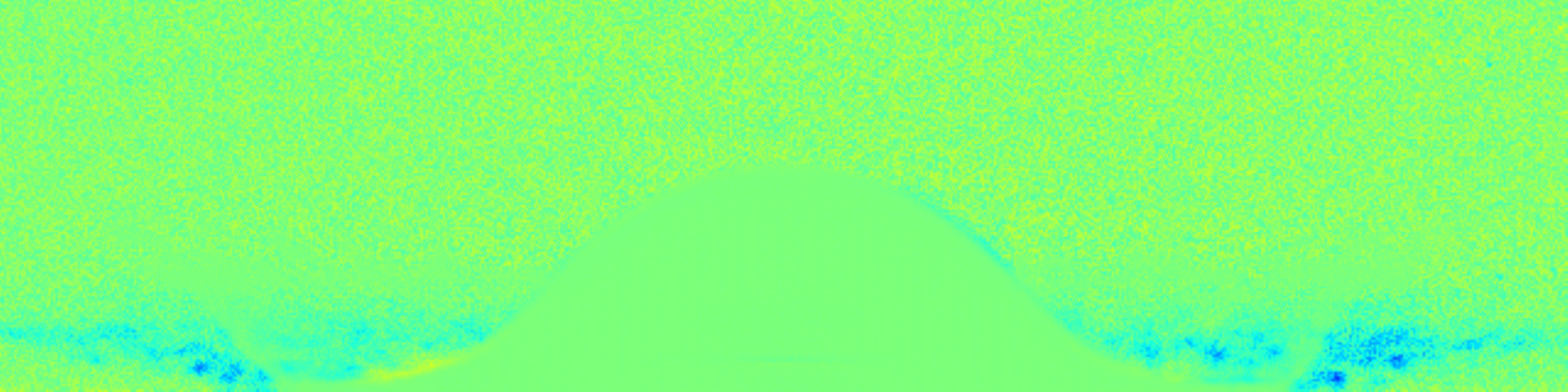}
\label{fig:251_Ws} }\\
\subfloat[]{
\includegraphics[width=0.9\columnwidth]{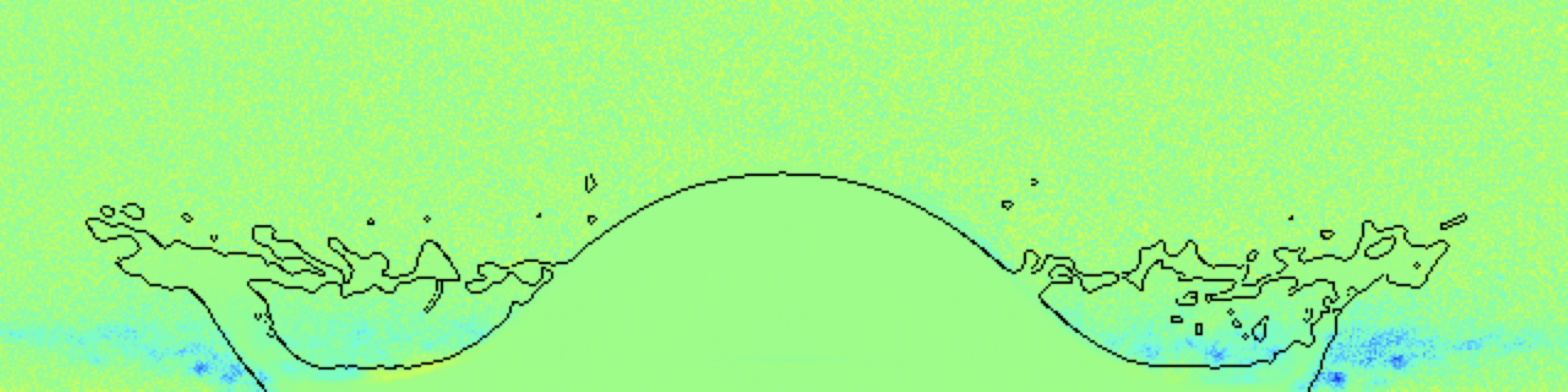}
\label{fig:251_Ws_contour}}
\caption{\label{fig:251_ori_Ws}
Typical splashing drop ($H = 60$~cm of Fig.~\ref{fig:cf}). 
\protect\subref*{fig:251_ori} Cropped image $\mathbf{x}$.
\protect\subref*{fig:251_Ws} Colormap of reshaped ${\bf Q}_2$.
\protect\subref*{fig:251_Ws_contour} Reshaped ${\bf Q}_2$ with black lines showing the drop's contour.
The blue-green-red (BGR) scale for \protect\subref*{fig:251_Ws} and \protect\subref*{fig:251_Ws_contour} is from $-0.040$ to $0.040$.}
\end{figure}
%===============================================

\begin{figure}[!ht]
\subfloat[]{
\includegraphics[width=0.9\columnwidth]{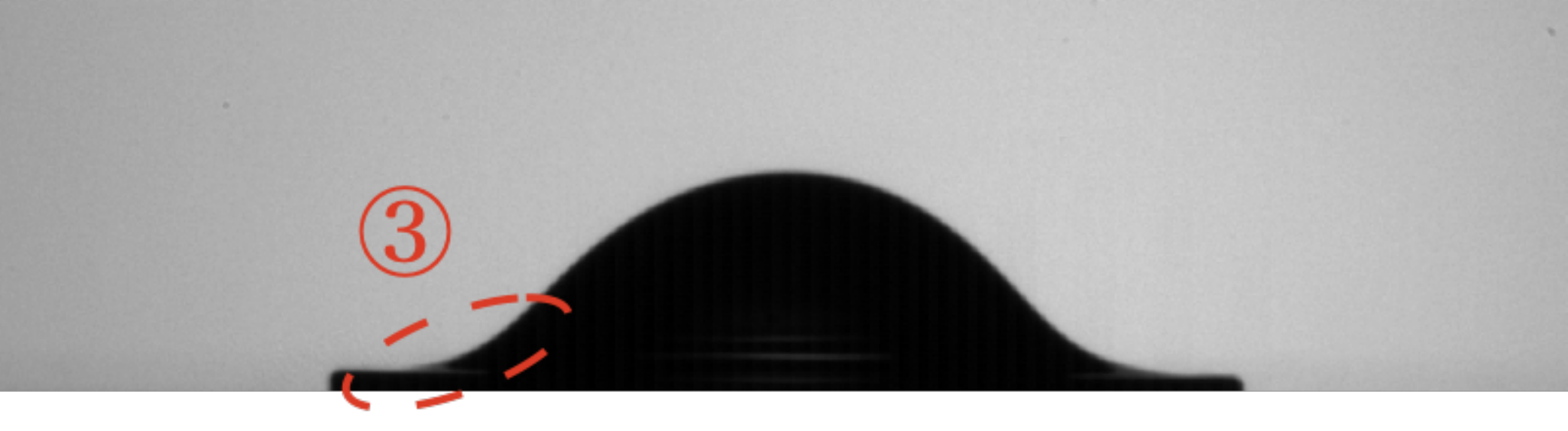}
\label{fig:32_ori} }
\\
\subfloat[]{
\includegraphics[width=0.9\columnwidth]{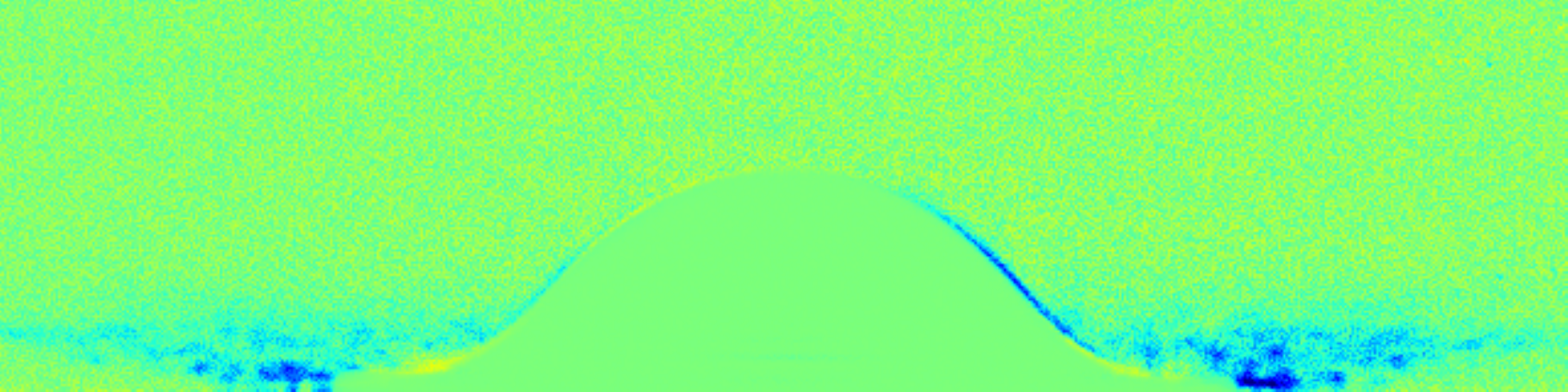}
\label{fig:32_Ws} }\\
\subfloat[]{
\includegraphics[width=0.9\columnwidth]{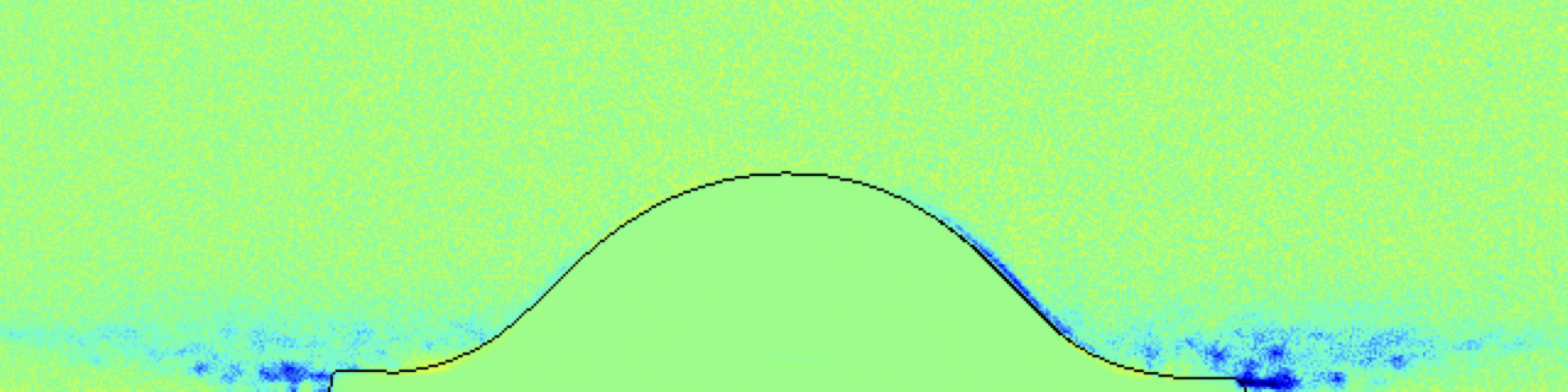}
\label{fig:32_Ws_contour} }
\caption{\label{fig:32_ori_Ws}
Typical nonsplashing drop ($H = 8$~cm of Fig.~\ref{fig:cf}). 
\protect\subref*{fig:32_ori} Cropped image $\mathbf{x}$.
\protect\subref*{fig:32_Ws} Colormap of the reshaped ${\bf Q}_2$.
\protect\subref*{fig:32_Ws_contour} Reshaped ${\bf Q}_2$ with black lines showing the drop's contour.
The blue-green-red (BGR) scale for \protect\subref*{fig:32_Ws} and \protect\subref*{fig:32_Ws_contour} is from $-0.040$ to $0.040$.
}
\end{figure}
%===============================================

%===============================================
\begin{figure}[!ht]
\centering
\includegraphics[width=\columnwidth]{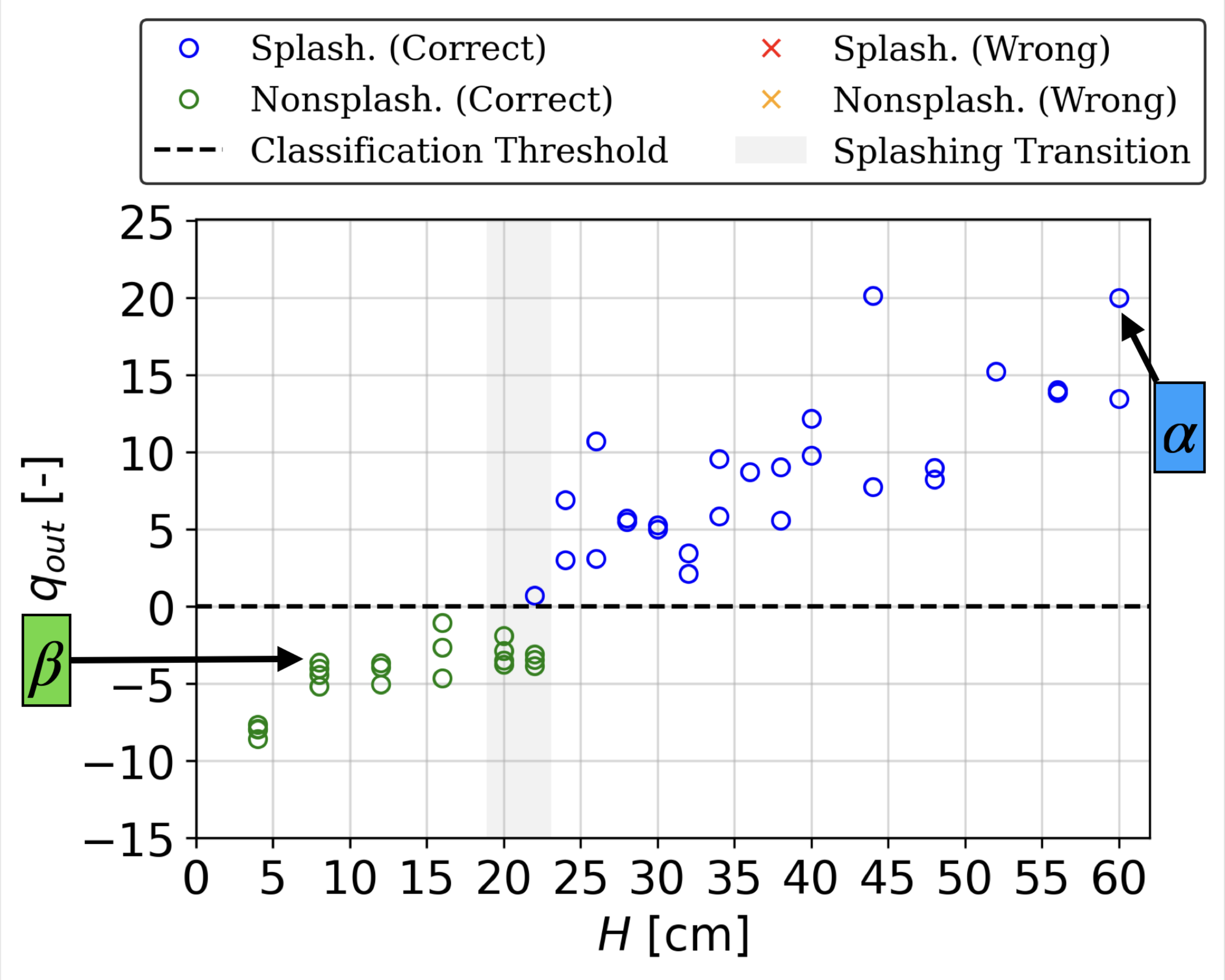}
\caption{\label{fig:qout_H_comb1} Plot of $q_{\mathrm{out},2}$ against impact height $H$ (combination 1).}
\end{figure}
%===============================================

Testing is the evaluation of the ability of the trained FNN to predict new images.
The results for all data combinations are shown in Table~\ref{tab:test_result}.
Among all combinations, the test accuracy in classifying images of both splashing and nonsplashing drops is $\geq94\%$.

To check how confident the trained FNN is with regard to classification, the splash probability $y_{\mathrm{pred},2}$ predicted by the FNN trained with combination 1 for test images with different impact heights $H$ is plotted in Fig.~\ref{fig:spl_prob}.
Note that only the plot for combination 1 is shown here, because similar results were obtained for other combinations.
For most splashing and nonsplashing drops, $y_{\mathrm{pred},2}$ is above 0.8 and below 0.2, respectively.
This indicates a reasonably high confidence of the trained FNN in classifying test images of both splashing and nonsplashing drops.

\subsection{\label{sec:visual}Extraction of image features of splashing and nonsplashing drops}

To extract the image features that the trained FNN identifies for classification, the important pixel positions were determined by reshaping and visualizing the trained weight matrix ${\bf W}$ as follows.
The matrix form of ${\mathbf W} \in \mathbb{R}^{C \times M}$ is
%===============================================
\begin{equation}
{\mathbf W} = 
\begin{bmatrix}
W_{1,1} &W_{1,2} &\dots &W_{1,M}\\
W_{2,1} &W_{2,2} &\dots &W_{2,M}
\end{bmatrix}
=
\begin{bmatrix}
{\mathbf w}_{1}\\
{\mathbf w}_{2}
\end{bmatrix}
\label{eq:W_h},
\end{equation}
%===============================================
where the elements ${\mathbf w}_{1}$ and ${\mathbf w}_{2}$ are the vector elements of the weight matrix for computing the probabilities of nonsplashing and splashing, respectively.
For visualization, both ${\mathbf w}_{1}$ and ${\mathbf w}_{2}$ were reshaped in row-major order into a two-dimensional matrix of size $h_\mathrm{img} \times w_\mathrm{img}$, which is the same shape as the input images, and are presented as colormaps in Fig.~\ref{fig:trained_w}.
In this figure, the combination column indicates the data combination that was used to train the FNN.
For both ${\mathbf w}_{1}$ and ${\mathbf w}_{2}$, the distribution of the values with large magnitude resembles a splashing drop.
These values are located at the same positions with opposite signs (negative blue and positive red).
The colormaps are similar for all data combinations, indicating good generalizability of the results.

In the colormap of the reshaped ${\mathbf w}_{2}$ of the FNN trained with each data combination, extreme negative values (blue) are distributed around \textcircled{1} and \textcircled{2}, while extreme positive values (red) are found around \textcircled{3}.
The distribution of theses values is symmetric.
Remarkably, by comparing these distribution with the images of a typical splashing drop [Fig.~\ref{fig:251_ori}] and a typical nonsplashing drop [Fig.~\ref{fig:32_ori}], it is found that \textcircled{1} corresponds to the area where the ejected secondary droplets of a splashing drop are present, \textcircled{2} to the contour of the main body of a splashing drop, and \textcircled{3} to the lamellae of a nonsplashing drop.

To understand how the trained weight $\bf W$ helps the FNN to classify images, the process of classifying the images of a typical splashing drop [Fig.~\ref{fig:251_ori}] and a typical nonsplashing drop [Fig.~\ref{fig:32_ori}] by the trained FNN is analyzed.
Note that these two typical images were cropped from Fig.~\ref{fig:cf} into size $h_\mathrm{img} \times w_\mathrm{img}$.

The analysis was done by visualizing ${\bf W}{\bf s}_\mathrm{in}$ from Eq.~(\ref{eq:lin_func_out}) as follows.
Each element of the trained weight matrix ${\bf W}$ was multiplied by the normalized intensity value at the corresponding pixel position of an image vector ${\bf s}_\mathrm{in}$ by computing the Hadamard product \cite{liu2008hadamard} ${\mathbf Q}_{i}$ using the following equation:
%===============================================
\begin{equation}
\mathbf{Q}_i = 
{\mathbf w}_i \circ {\mathbf{s}_\mathrm{in}^\top}
\label{eq:KP},
\end{equation}
%===============================================
where ${\mathbf w}_{i}$ is the vector element of the weight matrix for computing the probability of each output (splashing/nonsplashing) and $\mathbf{s}_\mathrm{in}^\top$ is the transpose of a flattened image.
Note that the sum of all elements of ${\mathbf Q}_i$ is equal to the matrix product of ${\bf w}_i$ and ${\bf s}_\mathrm{in}$, i.e., ${\bf w}_i{\bf s}_\mathrm{in}$.
For brevity, the explanation is focused on $\mathbf{Q}_{2}$, which corresponds to the output for splashing.
Similar to the visualization of $\mathbf{w}_{i}$, $\mathbf{Q}_{2}$ was reshaped in row-major order into a two-dimensional matrix of size $h_\mathrm{img} \times w_\mathrm{img}$ and presented as colormaps in Figs.~\ref{fig:251_Ws} and~\ref{fig:32_Ws}, with the same blue-green-red (BGR) scale as Fig.~\ref{fig:trained_w}, i.e., from $-0.040$ to $0.040$.

In Figs.~\ref{fig:251_ori} and~\ref{fig:32_ori}, with background lighting, the intensity value is almost zero at the pixel positions where the drop and the ejected secondary droplets covered the light.
Thus, in Figs.~\ref{fig:251_Ws} and~\ref{fig:32_Ws}, most of the values with large magnitudes (red and blue) of ${\bf w}_{2}$ were zeroed out (green) by the presence of the drop in both images.
This can be clearly observed in Figs.~\ref{fig:251_Ws_contour} and~\ref{fig:32_Ws_contour}, where there is only zero or green in the area bounded by the contours of the respective impacting drops.

%===============================================
\begin{table}[!t]
\caption{\label{tab:bias_out} Elements of the trained bias $\bf b$ of FNN trained with each data combination.}
\centering
\begin{tabular}{ccc}
\hline\hline
Combination & $b_2$ (Splashing) & $b_1$ (Nonsplashing)\\ \hline
1 & 0.0042 & $-0.0042$ \\ 
2 & 0.0035 & $-0.0035$ \\ 
3 & 0.0030 & $-0.0030$ \\ 
4 & 0.0042 & $-0.0041$ \\ 
5 & 0.0039 & $-0.0039$ \\
\hline\hline
\end{tabular}
\end{table}
%===============================================

Nevertheless, through careful observation at \textcircled{1} (the area where the ejected secondary droplets from a splashing drop are present) and \textcircled{2} (along the contour of the impacting drop) as indicated in Fig.~\ref{fig:trained_w}, there are more negative values (blue) remaining in the image of the nonsplashing drop than in that of the splashing drop.
As a consequence, the sum of all elements of ${\mathbf Q}_2$, i.e., ${\bf w}_{2}{\bf s}_\mathrm{in}$ [$\approx {q}_{\mathrm{out},2}$; see Eq.~(\ref{eq:lin_func_out})], for the image of the nonsplashing drop is lower than for the image of the splashing drop.
Thus, the image of a nonsplashing drop could not produce a value of $q_{\mathrm{out},2}$ that is high enough to exceed the classification threshold $y_\mathrm{pred} \geq 0.5$ to be classified as an image of a splashing drop.
On the other hand, for the image of the splashing drop, more negative values (blue) were zeroed out, raising the value of ${\bf w}_{2}{\bf s}_\mathrm{in}$.
Consequently, the value of $q_{\mathrm{out},2}$ is high enough to exceed the classification threshold $y_\mathrm{pred} \geq 0.5$ to be classified as an image of a splashing drop.
Note that owing to the presence of the sigmoid function, $y_\mathrm{pred} = 0.5$ when $q_\mathrm{out} = 0$.

The validity of these observations is confirmed by the plot of $q_{\mathrm{out},2}$ against impact height $H$ in Fig.~\ref{fig:qout_H_comb1}.
Since the same tendency was observed for all data combinations, only the plot for combination 1 is shown.
In this figure, the values of $q_{\mathrm{out},2}$ for the images of the splashing and the nonsplashing drops in Figs.~\ref{fig:251_ori} and~\ref{fig:32_ori} are indicated by $\alpha$ in the blue box and $\beta$ in the green box, respectively.
It is worth noting that $q_{\mathrm{out},2}$ shows an increasing trend with $H$.
Such a trend indicates that even without explicit learning, the trained FNN could estimate the inertia of an impacting drop based on the extracted image features, suggesting a possible correlation between the extracted image features and physical properties.

For the trained bias $\bf b$, the values for each data combination are listed in Table~\ref{tab:bias_out}.
The order of magnitude of $\bf b$ is $10^{-3}$, which is much smaller than the ${\bf q}_\mathrm{out}$ computed by the trained FNNs.
On the other hand, among the test images of all data combinations, the smallest absolute value of ${q}_\mathrm{out}$ is 0.40.
Therefore, ${\bf q}_\mathrm{out} \approx {\bf W}{\bf s}_\mathrm{in}$.
This indicates that the trained $\bf b$ did not affect the classification of the FNN and is negligible.

\subsection{\label{sec:class_criteria}Discussion of extracted image features}

In this subsection, the distribution of the values with large magnitude in the trained weight for splashing output $w_2$ is discussed in attempt to understand the underlying physical mechanism.

As shown in Fig.~\ref{fig:trained_w}, extreme negative values (blue) are distributed at \textcircled{1}, the area where the ejected secondary droplets of the splashing drop are present, and \textcircled{2}, along the contour of the main body of the impacting drop, while extreme positive values (red) are found at \textcircled{3}, the lamellae.
Among these, \textcircled{1} and \textcircled{2} are important characteristics for the FNN to identify a splashing drop, while \textcircled{3} is an important feature for the FNN to identify a nonsplashing drop.

The physical interpretations of \textcircled{1} and \textcircled{3} are quite intuitive.
In the case of \textcircled{1}, the physical interpretation is immediately obvious, since this feature satisfies the typical definition of a splashing drop, namely, the ejection of secondary droplets from the main body of the impacting drop \cite{josserand2016drop,yarin2006drop}.
In the case of \textcircled{3}, which is characteristic of a nonsplashing drop, it can be seen that the lamellae of a nonsplashing drop are shorter and thicker when $z_0/2R_{0} = 0.5$ compared with those of a splashing drop.
The lamellae are shorter because the ejection velocity of a lamella of a nonsplashing drop is lower owing to the lower impact velocity $U_{0}$ (smaller Weber number $We$) \cite{riboux2014experiments,riboux2017boundary}.
They are thicker because secondary droplets are not ejected from the lamellae of a nonsplashing drop.

\begin{figure*}[!ht]
\subfloat[]{
\includegraphics[width=\columnwidth]{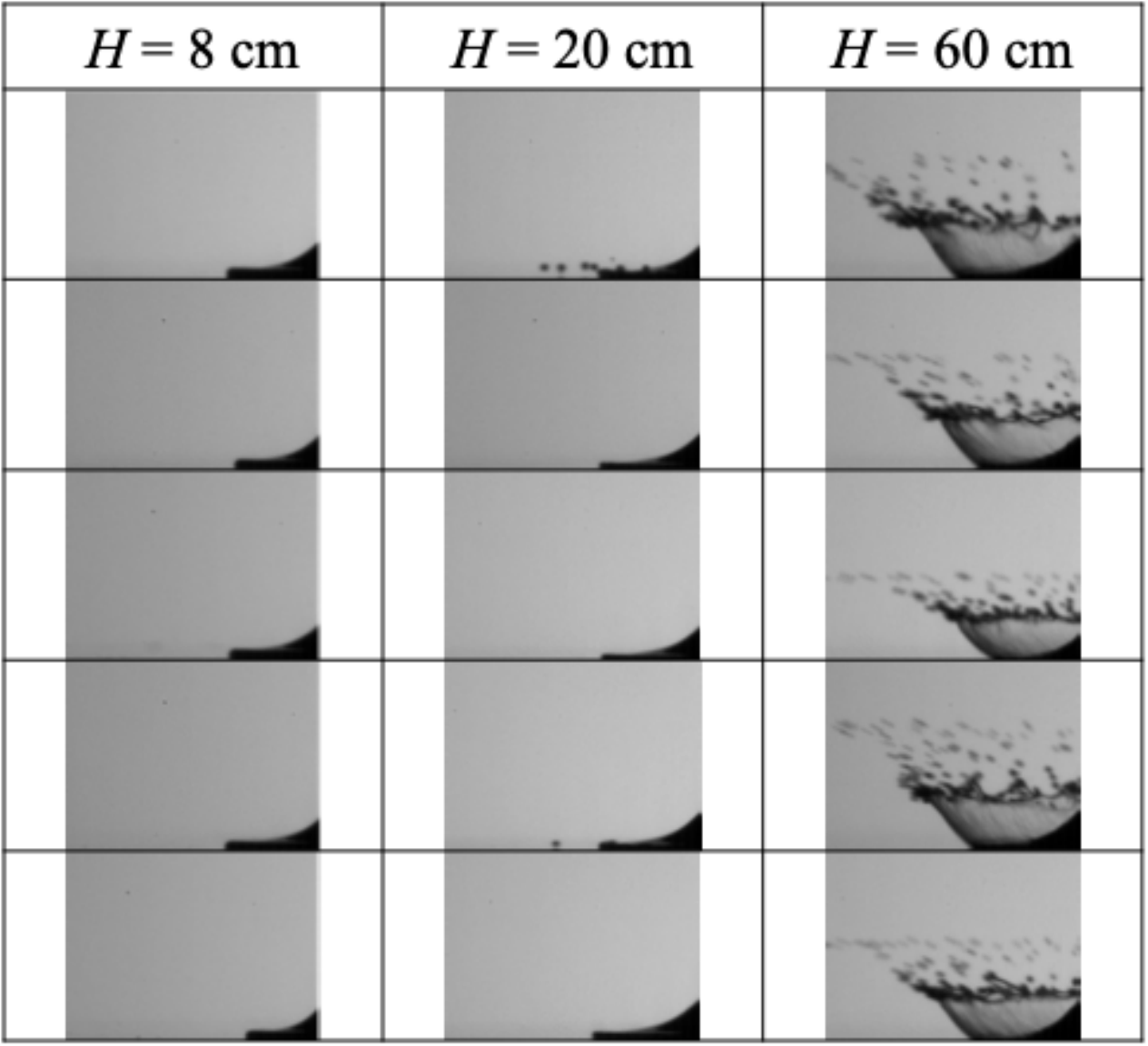}
\label{fig:img_lam}}
\hfill
\subfloat[]{
\includegraphics[width=\columnwidth]{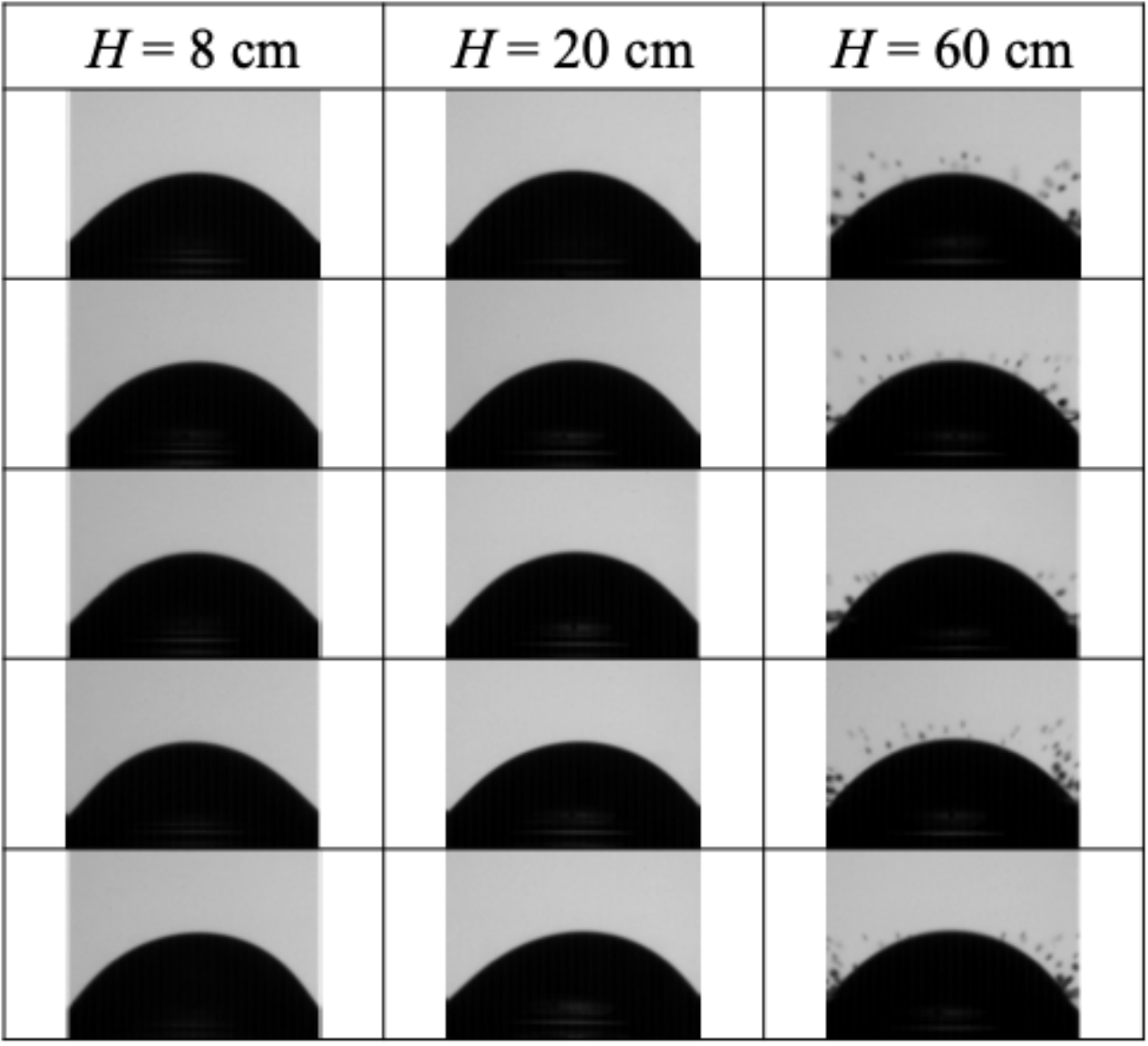}
\label{fig:img_mb} }
\caption{\label{fig:img_lam_mb}
Several examples of the images for the drop impacts with impact height $H = 8$, 20, and 60~cm, which were further cropped to focus on different parts of the impacting drop: \protect\subref*{fig:img_lam} the left lamella ($-1.8\geq r/2R_{0}\geq-0.6$); \protect\subref*{fig:img_mb} the main body ($-0.6\geq r/2R_{0}\geq0.6$).
}
\end{figure*}

\begin{figure*}[!ht]
\subfloat[]{
\includegraphics[width=\columnwidth]{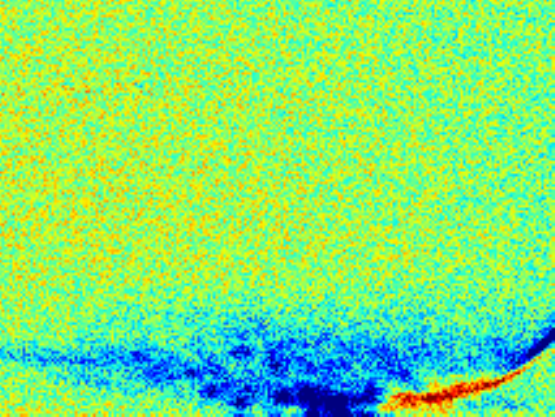}
\label{fig:w2_0.5D_lamella}}
\hfill
\subfloat[]{
\includegraphics[width=\columnwidth]{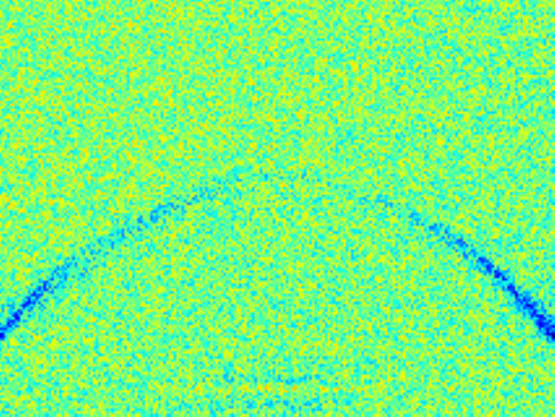}
\label{fig:w2_0.5D_contour} }
\caption{\label{fig:w2_0.5D}
Colormaps of reshaped ${\mathbf w}_{2}$ of the FNN trained with images of combination 1, which were further cropped to focus on different parts of the impacting drop:
\protect\subref*{fig:w2_0.5D_lamella} the left lamella ($-1.8\geq r/2R_{0}\geq-0.6$);
\protect\subref*{fig:w2_0.5D_contour} the main body ($-0.6\geq r/2R_{0}\geq0.6$).
The blue-green-red (BGR) scale is from $-0.040$ to $0.040$.
}
\end{figure*}

For the remainder of this subsection, the discussion will be focused on \textcircled{2}, the newly discovered characteristic of a splashing drop, which shows that the contour of the main body of a splashing drop is higher than that of a nonsplashing drop.
To understand how important \textcircled{2} is for the classification of splashing and nonsplashing drops, images extracted when $z_0/2R_{0} = 0.5$ were further cropped into two different sets: one focused on the left lamella ($-1.8\geq r/2R_{0}\geq-0.6$) and the other focused on the contour of the main body ($-0.6\geq r/2R_{0}\geq0.6$), where $r$ is the radial distance from the center of the drop.
Several examples of these two sets of cropped images are shown in Fig.~\ref{fig:img_lam_mb}.
These sets were used to train an FNN with the same architecture, and the trained weights were visualized as colormaps, which are shown in Fig.~\ref{fig:w2_0.5D}.
Note that these colormaps are scaled from $-0.040$ to $0.040$.
All the image features that were previously mentioned can be observed in these trained weights, where \textcircled{1} and \textcircled{3} can be seen in Fig.~\ref{fig:w2_0.5D_lamella}, while \textcircled{2} appears in Fig.~\ref{fig:w2_0.5D_contour}.
The respective test results are shown in Tables~\ref{tab:test_result_lamella} and~\ref{tab:test_result_contour}.
The accuracy of the FNN trained with the images focused on the lamella dropped slightly to $\geq 92\%$ for the classification of images of both splashing and nonsplashing drops for all combinations, as compared with the FNN trained with the images that have both the lamellae and the main body of the drop ($\geq94\%$).
Remarkably, the accuracy of the FNN trained with the images focused on the contour of the main body is still as high as $\geq78\%$ for the classification of images of both splashing and nonsplashing drops for all combinations, even without identifying the presence of the ejected secondary droplets.
In other words, an important but nonintuitive characteristic that differentiates a splashing drop and a nonsplashing drop has been successfully extracted through visualizing the image classification process of an FNN.
%===============================================
%===============================================

%===============================================
%===============================================

%===============================================
\begin{table*}[!t]
\caption{\label{tab:test_result_lamella} Test accuracy of FNN trained to classify splashing and nonsplashing drops through the left lamella ($-1.8\geq r/2R_{0}\geq-0.6$).}
\centering
\begin{tabular}{ccccccc}
\hline\hline
\multirow{2}{*}{Combination}& 
\multicolumn{6}{c}{Test accuracy} \\\cline{2-7} &
\multicolumn{2}{c}{Splashing} & \multicolumn{2}{c}{Nonsplashing} & \multicolumn{2}{c}{Total} \\ \hline
1& 27/27& 100\%& 21/21& 100\%& 
48/48& 100\% \\ 
2& 29/29& 100\%& 21/22& 95\%&
50/51& 98\% \\ 
3& 27/28& 96\%& 22/23& 96\%&
49/51& 96\% \\ 
4& 26/27& 96\%& 22/23& 96\%&
48/50& 96\% \\ 
5& 26/30& 87\%& 19/19& 100\%&
45/49& 92\% \\
\hline\hline
\end{tabular}
\end{table*}
%===============================================

%===============================================
\begin{table*}[!t]
\caption{\label{tab:test_result_contour} Test accuracy of FNN trained to classify splashing and nonsplashing drops through the main body ($-0.6\geq r/2R_{0}\geq0.6$).}
\centering
\begin{tabular}{ccccccc}
\hline\hline
\multirow{2}{*}{Combination}& 
\multicolumn{6}{c}{Test accuracy} \\\cline{2-7} &
\multicolumn{2}{c}{Splashing} & \multicolumn{2}{c}{Nonsplashing} & \multicolumn{2}{c}{Total} \\ \hline
1& 23/27& 85\%& 19/21& 90\%& 
42/48& 88\% \\ 
2& 24/29& 83\%& 20/22& 91\%&
44/51& 86\% \\ 
3& 21/28& 75\%& 21/23& 91\%&
42/51& 82\% \\ 
4& 25/27& 93\%& 17/23& 74\%&
42/50& 84\% \\ 
5& 22/30& 73\%& 16/19& 84\%&
38/49& 78\% \\
\hline\hline
\end{tabular}
\end{table*}
%===============================================

Understanding of the phenomenon of drop impact on a solid surface can be deepened through discovering the underlying mechanism that leads to \textcircled{2}, the higher contour of the main body of splashing drops when compared with nonsplashing drops.
Although the mechanism is unclear at the time of writing, it will be analyzed and discussed here from three aspects:
\begin{enumerate}
 \item[(i)] pre-impact drop shape;
 \item[(ii)] bubble entrainment;
 \item[(iii)] pressure impact.
\end{enumerate}

\begin{figure*}[!ht]
\subfloat[]{
\includegraphics[width=\columnwidth]{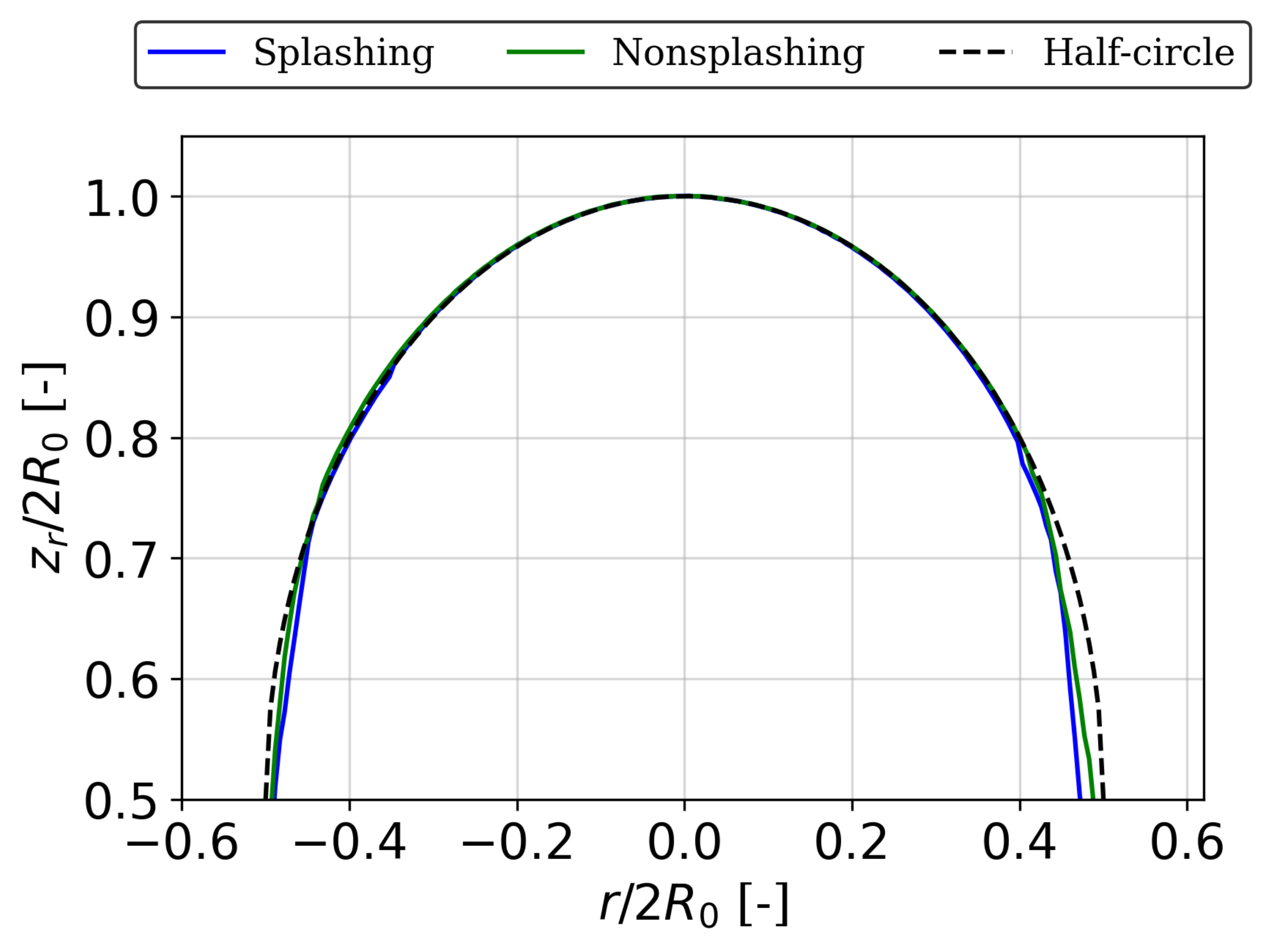}
\label{fig:outline_pre_impact}}
\hfill
\subfloat[]{
\includegraphics[width=\columnwidth]{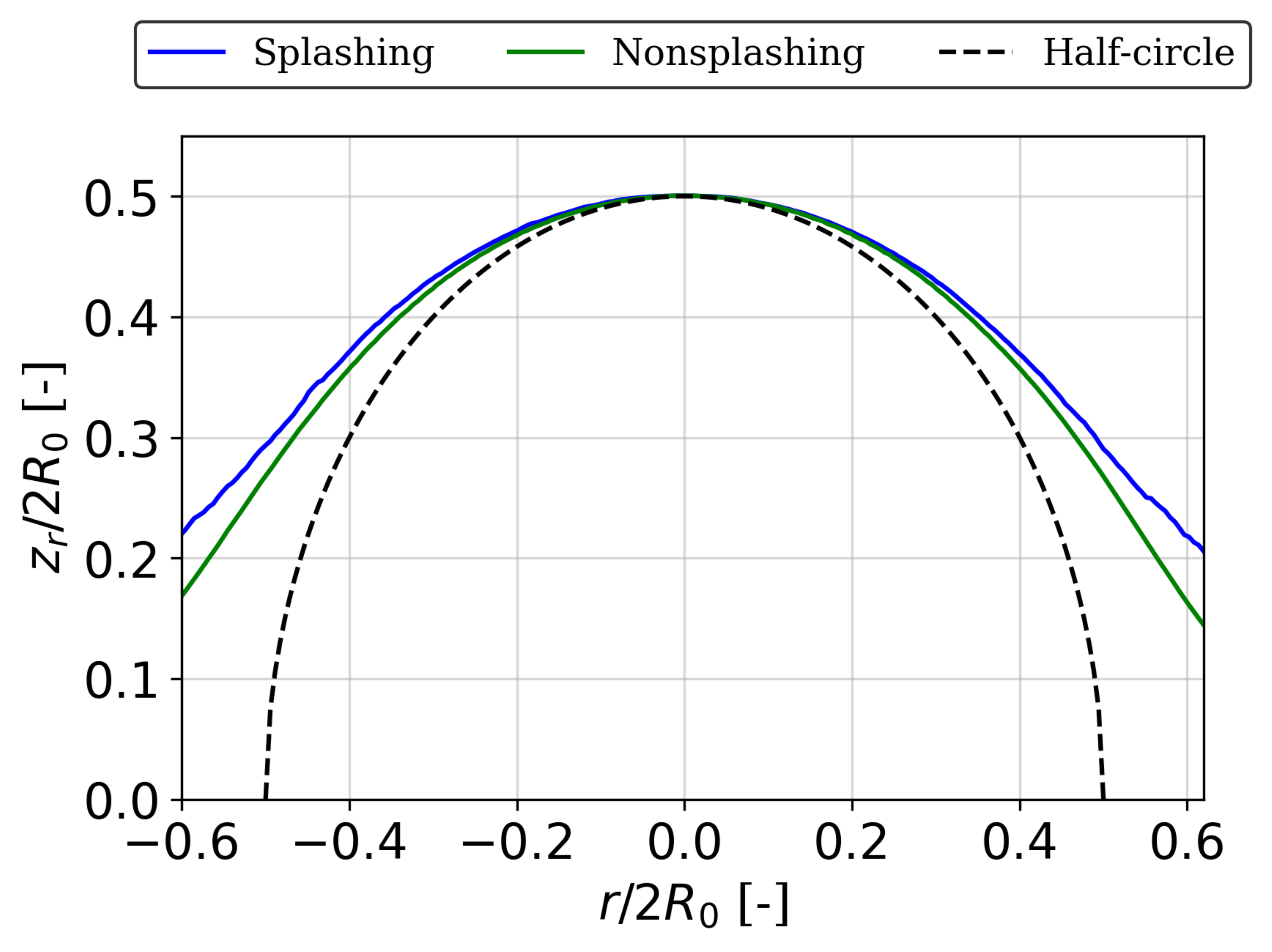}
\label{fig:outline_05D} }
\caption{\label{fig:outline} 
Averaged contours (normalized by area-equivalent diameter) of splashing and nonsplashing drops at different impact times: \protect\subref*{fig:outline_pre_impact} before impact ($h_c/2R_{0} \geq 1$); \protect\subref*{fig:outline_05D} when half of the drop impacted the surface ($h_c/2R_{0} = 0.5$).
}
\end{figure*}

(i) It is important to consider differences in pre-impact drop shape, because the difference in contour height could possibly be due to the difference in the pre-impact width-to-height ratio of the drop, which might have changed during free fall \cite{villermaux2009single}.
For the analysis, the contour of each drop before impacting the surface was extracted and averaged according to whether a drop was splashing or nonsplashing.
The contours, i.e., height of the drop along the radial axis $z_r$, were normalized by the area-equivalent diameter $2R_{0}$ and are shown in Fig.~\ref{fig:outline}, where the blue and green lines represent the averaged contours of splashing and nonsplashing drops, respectively.
The black dashed line represents the contour of a half-circle.
Before impact [see Fig.~\ref{fig:outline_pre_impact}], the averaged contours of both splashing and nonsplashing drops are similar to a half-circle, indicating that the averaged shapes of splashing and nonsplashing drops are similar.
However, after impact [see Fig.~\ref{fig:outline_05D}], the averaged contours of both splashing and nonsplashing drops become higher than the circle, with that of splashing drops being higher than that of nonsplashing drops.
This indicates that the higher contour of the main body of a splashing drop compared with that of a nonsplashing drop is due to the dynamics during the impact, rather than to any difference in drop shape before impact.

%===============================================
%===============================================
%===============================================
\begin{figure*}[!ht]
\centering
\subfloat[]{
\includegraphics[width=0.8\textwidth]{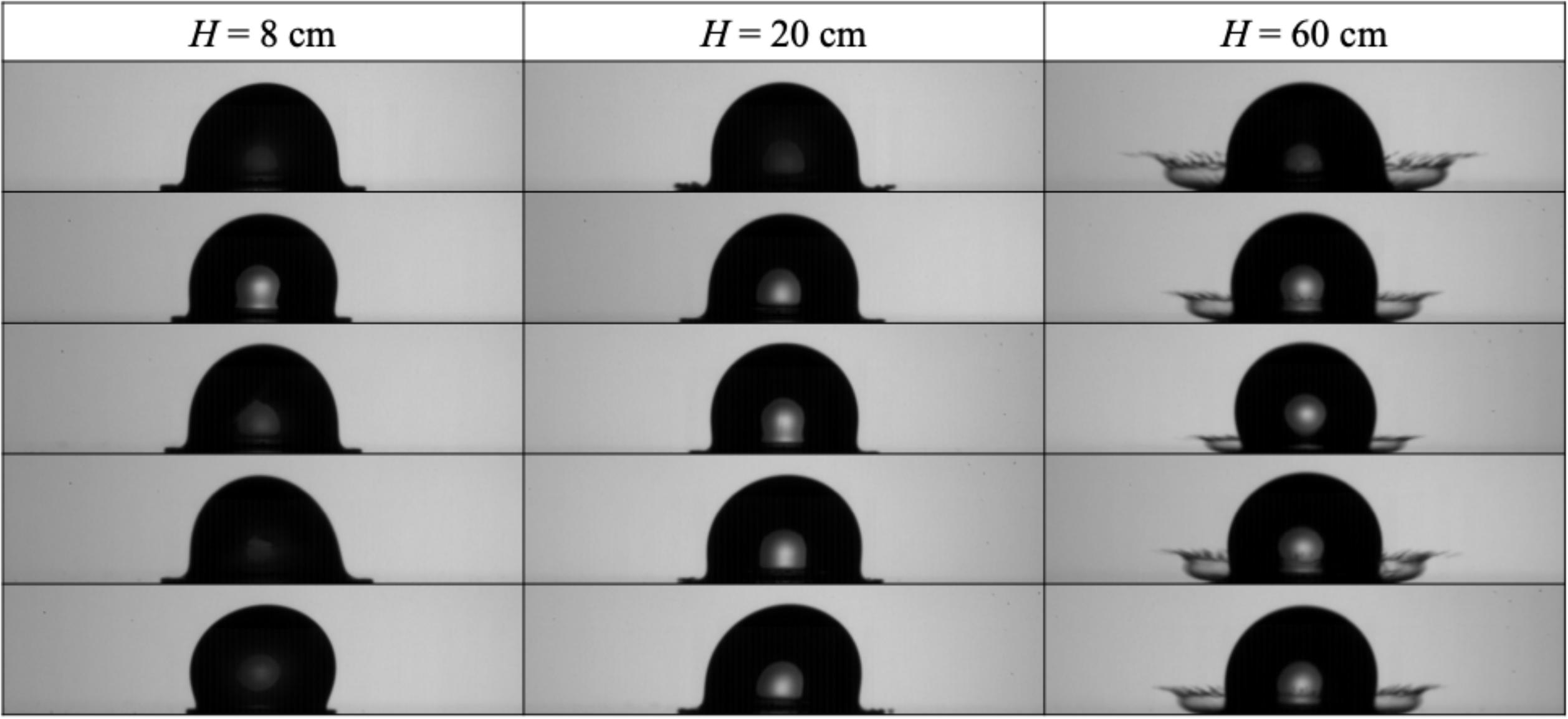}
\label{fig:img_0.75D}}
\hfill
\subfloat[]{
\includegraphics[width=0.8\textwidth]{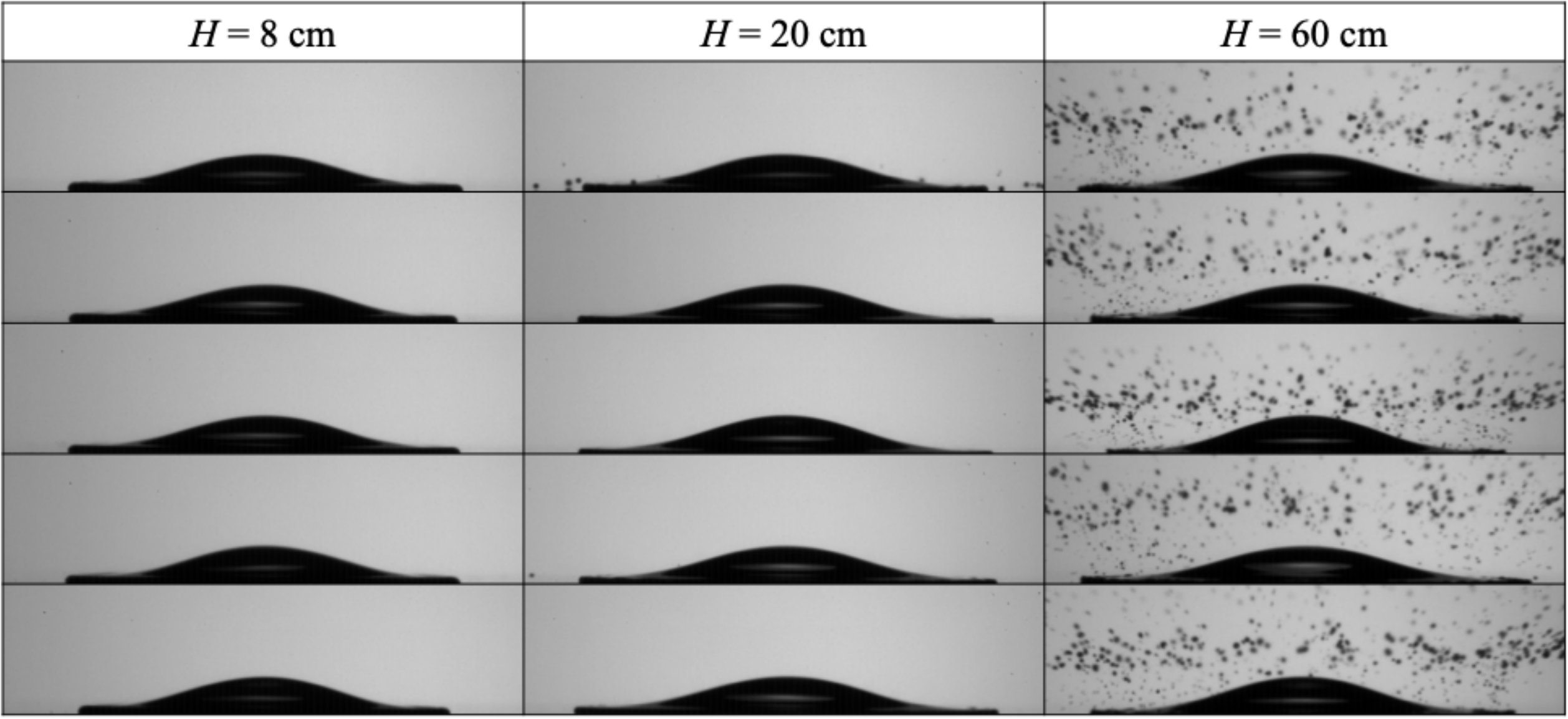}
\label{fig:img_0.25D} }
\caption{\label{fig:img_time} 
Several examples of the images for the drop impacts with impact height $H = 8$, 20, and 60~cm, extracted from different dynamical regimes: \protect\subref*{fig:img_0.75D} the pressure-impact regime $z_0/2R_{0} = 0.75$; \protect\subref*{fig:img_0.25D} the self-similar inertial regime $z_0/2R_{0} = 0.25$.
}
\end{figure*}
%===============================================

%===============================================
\begin{figure}[!ht]
\centering
\subfloat[]{
\includegraphics[width=\columnwidth]{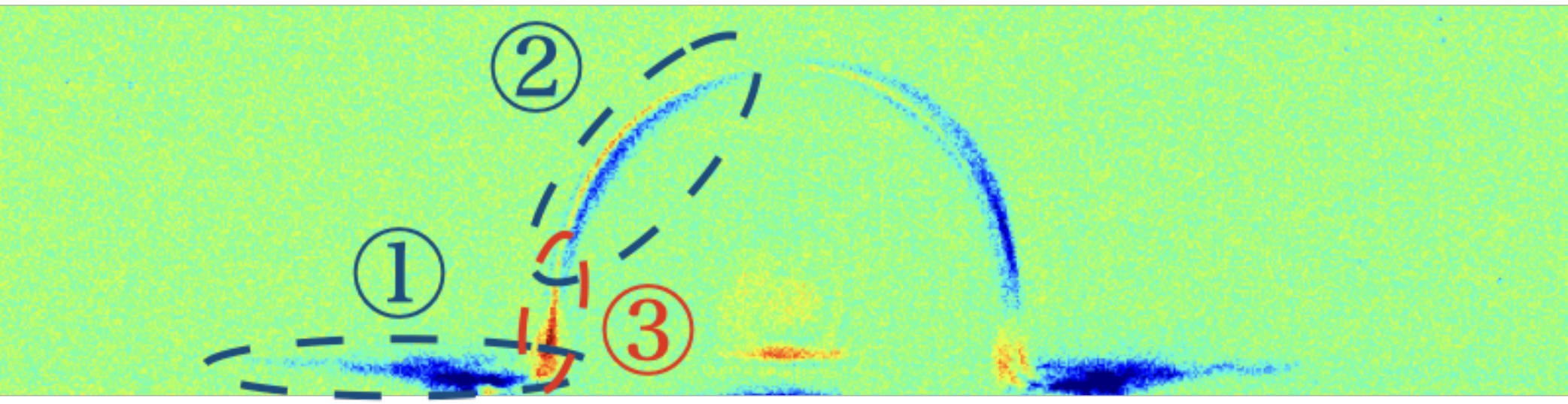}
\label{fig:w2_0.75D}}
\hfill
\subfloat[]{
\includegraphics[width=\columnwidth]{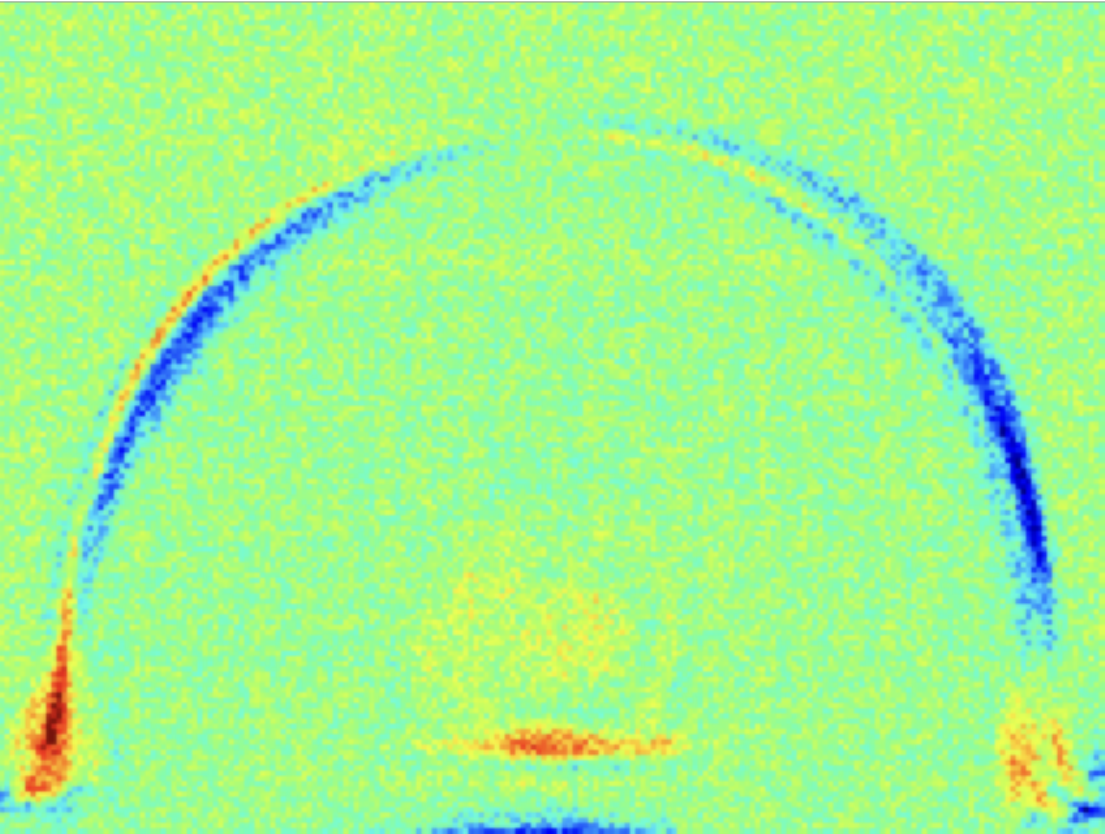}
\label{fig:w2_0.75D_contour}}
\hfill
\subfloat[]{
\includegraphics[width=\columnwidth]{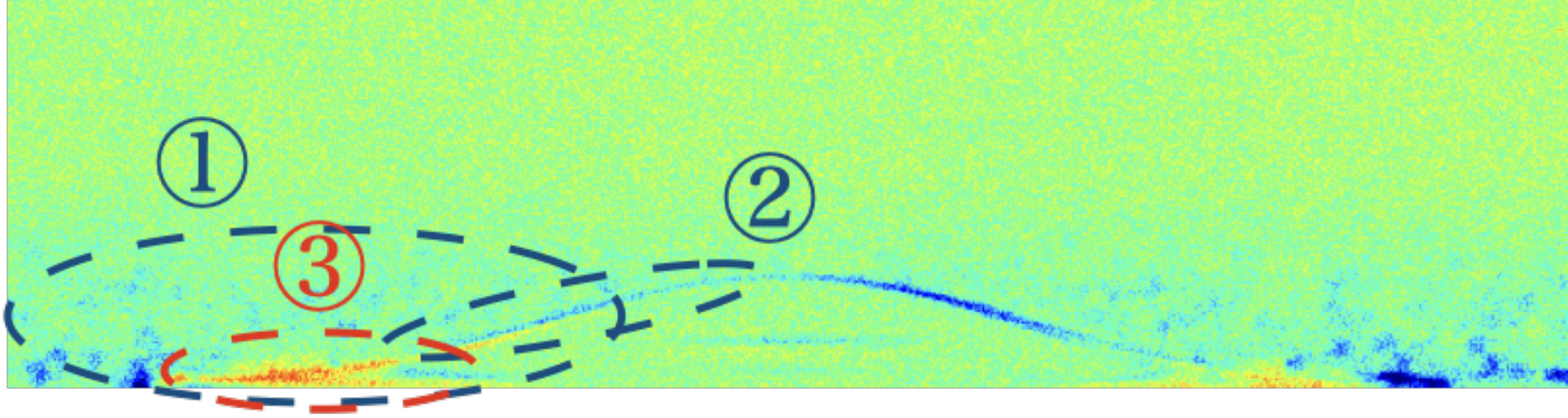}
\label{fig:w2_0.25D} }
\caption{\label{fig:w2_time} 
Colormaps of reshaped ${\mathbf w}_{2}$ of FNN trained with images of combination 1 in different dynamical regimes:
\protect\subref*{fig:w2_0.75D} the pressure-impact regime $z_0/2R_{0} = 0.75$ and \protect\subref*{fig:w2_0.75D_contour} the zoomed-in view;
\protect\subref*{fig:w2_0.25D} the self-similar inertial regime $z_0/2R_{0} = 0.25$.
The blue-green-red (BGR) scale is from $-0.040$ to $0.040$.
}
\end{figure}
%===============================================

(ii) Bubble entrainment is analyzed because the difference in contour height could possibly be due to the difference in the volume of air entrapped.
In the study by Bowhuis \emph{et al.} \cite{bouwhuis2012maximal}, experiments were conducted on drop impact for Stokes numbers $St$ ranging from $10^2$ to $10^5$.
The results showed that the volume of air entrapped during drop impact increases with $St$ owing to the reduction in capillary forces, until it reaches a maximum value at $St = 10^4$, after which it decreases with increasing $St$ owing to the increasing inertia of the drop.
Since splashing did not occur in the study by Bowhuis \emph{et al.}, their results validate those reported here, in which there was a splashing transition at $St \approx 10^5$.
However, all these results also prove that the higher contour of a splashing drop is not due to air entrapment, because the range of $St \geq 10^5$ is in the regime where air entrapment is inhibited by the inertia of the drop.

(iii) Last, but not least, the pressure impact is analyzed because the difference in contour height could be due to the reaction force of the pressure impact that arises when a drop collides with a solid surface.
From the studies by Eggers \emph{et al.} \cite{eggers2010drop} and Lagubeau \emph{et al.} \cite{lagubeau2012spreading}, it is known that the spreading dynamics experiences a transition between pressure-impact and self-similar inertial regimes when $z_0/2R_{0} = 0.5$.
The difference between splashing and nonsplashing drops in these two dynamical regimes were checked by training an FNN with the same architecture to classify images of splashing and nonsplashing drops in both regimes: $z_0/2R_{0} = 0.75$ (pressure-impact regime) and 0.25 (self-similar inertial regime).
Several examples of the images used for the training are shown in Fig.~\ref{fig:img_time}.
The trained weights were visualized as colormaps, scaled from $-0.040$ to $0.040$, which are shown in Fig.~\ref{fig:w2_time}.
In this figure, Fig.~\ref{fig:w2_0.75D_contour} shows the zoomed-in view of Fig.~\ref{fig:w2_0.75D}.
Interestingly, for both $z_0/2R_{0} = 0.75$ and 0.25, the extracted image features correspond to those extracted from the image classification for $z_0/2R_{0} = 0.5$.
The higher contour of the main body of a splashing drop compared with a nonsplashing drop can already be observed in the pressure impact regime $z_0/2R_{0} = 0.75$.
It is therefore necessary to examine the relationship between pressure impact and image feature \textcircled{2}.
It is known that pressure $P \propto \rho R_0 \dot{U}$, where $\dot{U}$ is the acceleration of the drop during impact.
Since $\rho$ and $R_0$ were fixed in the experiment, the double integral of $P$ with respect to time is equivalent to a length scale.
By checking whether this length scale scales with impact velocity $U_0$, the relationship between $P$ and the higher contour can be more or less confirmed.
For this, double integration with respect to time was performed on the expression for the dimensionless pressure exerted by an impacting drop on a solid surface proposed by Philippi \emph{et al.} \cite{philippi2016drop}, namely, $\hat{P} = 3/\pi \sqrt{3\hat{t} - \hat{r}^2}$, where $\hat{t}$ is the dimensionless time and $\hat{r}$ is the dimensionless radial distance from the center of the drop.
The result scales only with $\hat{t}$, i.e., $\iint \hat{P}\, dt \,dt \propto \hat{t}^{3/2}$.
Since $\hat{t}$ is the same for the same $z_0$, the contour height is expected to be the same for both splashing and nonsplashing drops.
This analysis cannot prove that pressure impact is the direct cause of the higher contour of a splashing drop compared with a nonsplashing that was found by the trained FNN.
However, this does not rule out the existence of a relationship between pressure impact and contour height.
Therefore, further analysis is necessary to clarify the mechanism underlying the difference in contour height between splashing and nonsplashing drops.

\section{\label{sec:conclusion}Conclusions and Outlook}

In this study, nonintuitive characteristics of a drop splashing on a solid surface have been unveiled through image feature extraction using a feedforward neural network (FNN).
Experiments were carried out to collect images of splashing and nonsplashing ethanol drops impacting on a hydrophilic glass surface after falling from an impact height $H$ ranging from 4 to 60~cm.
The collected images were processed to produce very similar images for the training, validation, and testing of the FNN.

The trained FNN achieved an accuracy $\geq 96\%$ during testing to classify images of splashing and nonsplashing drops when half of the drop impacted the surface ($z_0/2R_{0} = 0.5$).
The confidence of the FNN with regard to the classification was reasonably high, with a splashing probability $y_{\mathrm{pred},2}\geq 0.8$ and $\leq 0.2$ for most of the images of splashing drops and nonsplashing drops, respectively.

Analysis of the classification process showed that the important image features used by the trained FNN to identify a splashing drop are the area where the ejected secondary droplets are present and along the contour of the main body of the impacting drop, while the relevant features used to identify a nonsplashing drop are short and thick lamellae.
Among these features, the presence of ejected secondary droplets has typically been used in previous studies to distinguish splashing from nonsplashing drops, while short and thick lamellae have been identified as being characteristic of nonsplashing drops.
However, the higher contour of the main body of a splashing drop compared with a nonsplashing drop has not hitherto been reported.
Further image classification shows that the trained FNN has an accuracy $\geq 82\%$ in classifying splashing and nonsplashing drops according to the contour of the main body without checking for the presence of ejected secondary droplets.

Last, but not least, these image features quantified by $q_{\mathrm{out},2}$ exhibit an increasing trend with impact height $H$, indicating a correlation between image features and impact velocity.
This opens up the possibility of image-based estimation of impact velocity during drop impact.

Further experimental and computational studies are crucial for obtaining greater understanding of the mechanism responsible for the higher contour of the main body of a splashing drop.
Moreover, it is also important to discover time-dependent image features of a splashing drop, which can be done by training ANNs for classification based on high-speed videos of drop impact instead of just still images.

For the outlook of this study, through transfer learning \cite{weiss2016survey,zhuang2021comprehensive,inubushi2020transfer}, the FNNs trained in this study can be further trained using images that are collected from other drop impact experiments, which have other physical quantities as the manipulated variables.
Eventually, a universal splashing-nonsplashing classification model, which can unveil the nonintuitive universal characteristics of a splashing drop, can be built.
For that, the image data and the FNN coding used in this study would be uploaded on GitHub (the GitHub link is under construction) and we would like to invite other drop impact researchers to build this universal classification model together.

We believe that through the methodology of this study that utilizes the image processing ability of ANNs and visualizes the classification processes that are usually black-box processes, nonintuitive characteristics of various phenomena related to fluid dynamics can be extracted, thus creating new insights for the research of fluid dynamics.
Therefore, we would also like to extend our invitation to researchers who have collected numerous beautiful data (not just limited to images) of various phenomena, so that together we can develop the methodology and explore different fluid phenomena from a different perspective.

\section*{Acknowledgement}

This work was funded by Japan Society for the Promotion of Science (Grant No. 20H00223, 20H00222, and 20K20972) and Japan Science and Technology Agency PRESTO (Grant No. JPMJPR21O5).
The authors would also like to thank Dr. Masaharu Kameda (Professor, Tokyo University of Agriculture and Technology) and Dr. Masakazu Muto (Assistant Professor, Tokyo University of Agriculture and Technology) for their valuable discussions and suggestions.

\section*{Author Declarations}

The authors have no conflicts of interests to disclose.

\section*{Data availability}

The data that support the findings of this study are available from the corresponding authors upon reasonable request.

\section*{Appendix}

\appendix

\section{\label{app:arch_opt} Architecture Optimization}

The architecture of the feedforward neural network (FNN) was optimized to find the optimum numbers of hidden layers and neurons to achieve the desired performance.
The desired performance is set to be loss $\leq 0.30$ and accuracy $\geq 80\%$ for images of both splashing and nonsplashing drops.
To reduce the computational cost, the architecture with the least number of hidden layers that was still capable of achieving the desired performance was chosen as the optimized architecture.

For the training of all candidates, the training results were similar in terms of test accuracy and extracted image features.
In terms of test accuracy, all the candidates achieved an accuracy of $\geq$ 80\%.
With regard to the colormaps of the trained weight ${\mathbf w}_{i}$, the distribution of the values with large magnitude of the active neurons is similar for all candidates.
The optimized architecture that achieved the desired performance with the lowest computational cost has zero hidden layers.

\bibliographystyle{ieeetr}%elsarticle-num
\bibliography{ref}

\end{document}